\documentclass[twocolumn,aps,superscriptaddress,showpacs,preprintnumbers,amsmath,amssymb]{revtex4}
\usepackage{graphicx,color}
\usepackage{dcolumn}
\usepackage{colordvi} 
\usepackage{bm}
\begin{document}
\title{Cosmology with space-based gravitational-wave detectors\\ 
--- dark energy and primordial gravitational waves ---}
\author{Atsushi Nishizawa}
\email{anishi@yukawa.kyoto-u.ac.jp}
\affiliation{Yukawa Institute for Theoretical Physics, Kyoto University, 
Kyoto 606-8502, Japan}
\author{Kent Yagi}
\affiliation{Department of Physics, Kyoto University, 
Kyoto 606-8502, Japan}
\author{Atsushi~Taruya}
\affiliation{Research Center for the Early Universe, Graduate School of 
Science, The University of Tokyo, Tokyo 113-0033, Japan}
\affiliation{Institute for the Physics and Mathematics of the Universe, 
The University of Tokyo, Kashiwa, Chiba 277-8568, Japan}
\author{Takahiro Tanaka}
\affiliation{Yukawa Institute for Theoretical Physics, Kyoto University, 
Kyoto 606-8502, Japan}

\date{\today}

\begin{abstract}
Proposed space-based gravitational-wave (GW) detectors such as DECIGO and BBO 
will detect $\sim10^6$ neutron-star (NS) binaries and determine the luminosity distances to the binaries with high precision. Combining the luminosity distances with cosmologically-induced phase corrections on the GWs, cosmological expansion out to high redshift can be measured without the redshift determinations of host galaxies by electromagnetic observation and be a unique probe for dark energy. On the other hand, such a NS-binary foreground should be subtracted to detect primordial GWs produced during inflation. Thus, the constraining power on dark energy and the detectability of the primordial gravitational waves strongly depend on the detector sensitivity and are in close relation with one another. In this paper, we investigate the constraints on the equation of state of dark energy with future space-based GW detectors with/without identifying the redshifts of host galaxies. We also study the sensitivity to the primordial GWs, properly dealing with the residual of the NS-binary foreground. Based on the results, we discuss the detector sensitivity required to achieve the forementioned targeted study of cosmology.
\end{abstract}

\maketitle

\section{Introduction}

Future space-based gravitational-wave (GW) detectors such as DECI-hertz Interferometer Gravitational-wave Observatory (DECIGO) \cite{bib1,bib2} and Big-Bang Observer (BBO) \cite{bib3} (see also \cite{bib4} for updated information) are the most sensitive to GWs in $0.1 - 1\,{ \rm{Hz}}$ band and will aim at detecting the primordial GW background, the mergers of intermediate-mass black holes (BH), and a large number of neutron-star (NS) binaries in an inspiraling phase. These GW sources enable us to probe inflation and density perturbation in the early universe \cite{bib46,bib70,bib67,bib68,bib69}, to measure the cosmic expansion with unprecedented precision \cite{bib4,bib37}, to investigate the population and formation history of compact binary objects \cite{bib64}, and to test alternative theories of gravity \cite{bib5,bib65,bib66,bib48}. Therefore, DECIGO and BBO will open up the window of gravitational-wave cosmology.

It is known that the continuous GW signal from a compact-binary object provides a unique way to measure the luminosity distance to the source with high precision. Such binary sources are often referred to as the standard siren (analogous to the electromagnetic standard candle). With the redshift information determined by an electromagnetic follow-up observation, the standard siren can be an accurate tracer of the cosmic expansion \cite{bib6}. The potential power of this method as a dark-energy probe has been investigated with ground- and space-based detector configurations for GWs \cite{bib4,bib21,bib22,bib23,bib24,bib25,bib26,bib27,bib28, bib88}. Although these detectors will constrain the evolution of dark-energy, most of the preceding works assume that the redshifts of all GW sources are known by electromagnetic observations. This assumption is rather strong and is too optimistic to be justified, because the spectroscopic follow-up observation of galaxies is, in general, time-consuming, particularly at high redshifts \cite{bib84}. In addition, not all galaxies can be observed due to intrinsic faintness, the absence of spectral features, limited sky coverage, and a limited redshift range (redshift desert). As a result, in a practical follow-up observation, the fraction of the binary sources whose redshifts are spectroscopically obtained is significantly reduced. From a rough estimate based on the number density of galaxies potentially observable and the number of galaxies to be observed in the future galaxy redshift surveys, it turns out that the fraction of redshift identification could be $\sim 10^{-4}$ with uncertainty of about one order of magnitude (See Sec.~\ref{sec5E} in detail). This means that the GW sources with redshift information would be rare unless we perform a large-scale follow-up campaign dedicated for GW events. Thus, this reduces the number of binary sources available as a standard siren. 

Another approach is to measure the cosmic-expansion rate from GW observations alone without electromagnetically-estimated redshifts. As suggested in \cite{bib1}, the cosmic expansion affects not only the amplitude (luminosity distance) of GWs but also the phase, and it is possible to directly obtain information about the cosmic acceleration by accurately measuring the GW-phase shift of a binary source at a certain redshift. Although the redshifts of the binaries are assumed to be determined by the electromagnetic follow-up observations in the previous works \cite{bib1,bib20}, we show in this paper that we can measure the expansion history of the universe without any reference to the electromagnetic counterpart or host-galaxy identification by combining the luminosity distance and the cosmological phase shift, which are both independently determined from the amplitude and phase of GWs. This method based on a purely GW observation enables us to compare the observational data with those obtained in other electromagnetic observations. Since proposed space-based GW detectors such as DECIGO and BBO would detect $\sim10^6$ NS binaries, they provide a novel opportunity to measure the property of dark energy without using the cosmic ladders \cite{bib4}. 

Our primary interest here is in the potential of the standard siren without any help from redshift information of the binary sources. However, we also show how the sensitivity is improved by adding the redshift information of a fraction of NS binaries. As a result, we found that the contribution of binary sources whose redshift is determined tightly constrain the dark-energy parameters if the fraction of redshift identification is larger than $10^{-5}$ - $10^{-4}$. In some cases that the fraction of redshift identification by the future galaxy redshift survey is smaller than the one we expect, the contributions of binary sources with/without redshifts are comparable. Therefore, our study in this paper indicates that the GW standard siren without redshift information guarantees minimally achievable constraint on the dark energy and that the follow-up observation of the host galaxies targeted at the GW sources is crucial to improve the sensitivity.   

Another important scientific target of DECIGO and BBO is a primordial GW background. For its search, one needs to accurately identify the waveforms of NS binaries and subtract them up to the level sufficiently below the amplitude of the primordial GW background \cite{bib15,bib17}. Thus, an accurate determination of the waveforms of NS binaries is essential both for the utilization of the NS binaries as the standard siren and the detection of the primordial GW background. For this reason, in this paper, we comprehensively treat these issues and investigate the scientific outcome obtained by DECIGO and BBO, allowing their noise curves to be scaled appropriately and including confusion noises produced by astrophysical sources. 
The scaling of the noise curve is worth considering, because the requirement for instrumental noise is flexible at the preconceptual stage of the detector configuration and should be determined to optimize the scientific results. 

This paper is organized as follows. In Sec.~\ref{sec2}, we briefly review the detector configurations of DECIGO and BBO and provide a model for the power spectra of detector noises and astrophysical confusion noises that we use throughout this paper. In Sec.~\ref{sec3}, we present the model of a GW waveform from a compact binary and give an estimate of the signal-to-noise ratio obtained when a matched-filtering analysis is performed. In Sec.~\ref{sec4}, we explain the procedure of the NS-binary subtraction and give the residual contribution to the total noise after the subtraction. Based on these results, we calculate the detector sensitivities to dark energy in Sec.~\ref{sec5} and to a primordial GW background in Sec.~\ref{sec6}. In Sec.~\ref{sec7} we give discussions on the results and the feasibility of each method. Sec.~\ref{sec8} is devoted to conclusions. Throughout the paper, we adopt units 
$c=G=1$.

\section{Detector noise and astrophysical foregrounds}
\label{sec2}
In this section, we summarize the detector configuration of DECIGO and BBO and provide detector-noise curves that are used throughout the paper.

\subsection{DECIGO/BBO}

In the current preconceptual design of the detector \cite{bib2,bib3,bib4}, DECIGO and BBO orbit the Sun with a period of one sidereal year, and constitute four clusters, each of which consists of three spacecrafts exchanging laser beams with the others. Two of the four clusters are located at the same position to enhance the correlation and hence the sensitivity to a stochastic GW background, and the other two are widely separated from each other on the Earth orbit in order to enhance the angular resolution so that we can easily identify the host galaxy of each NS binary via the electromagnetic follow-up observations \cite{bib4}.

DECIGO and BBO are the most sensitive to GWs in the $0.1 - 1\,{\rm{Hz}}$ band, which is determined so as to avoid the astrophysical GW foreground produced by white-dwarf (WD) binaries. DECIGO instrumental noise is dominated by radiation pressure noise below $\sim 0.2\,{\rm{Hz}}$ and by laser shot noise above that frequency. As for BBO, the sensitivity at high frequencies is limited by beam-pointing-jitter and stray light noises rather than laser shot noise. The noise curve of DECIGO is calculated adopting currently-proposed design parameters \cite{bib2,bib11}. We obtained fitting formula for the sky-averaged noise curve of DECIGO single interferometer: 
\begin{align}
\hat{S}_{\rm{h,\,D}}^{\rm{inst}} (f) &= 3.30\times 10^{-50} \biggl(\frac{f}{1\rm{Hz}} \biggr)^{-4} \nonumber \\
&+3.09\times 10^{-47} \left[1+\left(\frac{f}{f_{\rm{c}}}\right)^2 \right] \;\; 
\rm{Hz}^{-1} \;, 
\label{eq4}
\end{align}
where $f_{\rm{c}}=7.69\,{\rm{Hz}}$. On the other hand, BBO sky-averaged noise curve is given in \cite{bib4} as 
\begin{align}
\hat{S}_{\rm{h,\,B}}^{\rm{inst}} (f) &= 6.15\times 10^{-51} \biggl(\frac{f}{1\rm{Hz}} \biggr)^{-4} 
+1.95\times 10^{-48} \nonumber \\
&+1.20\times 10^{-48} \biggl(\frac{f}{1\rm{Hz}} \biggr)^{2}  \;\; 
\rm{Hz}^{-1} \;.
\label{eq5}
\end{align}

DECIGO and BBO have different noise shapes, resulting from the difference in the interferometer type and optical parameters. The default noise curve of DECIGO is designed to avoid WD confusion noise at low frequencies. On the other hand, BBO has slightly better sensitivity than DECIGO in the frequency range below several ${\rm{Hz}}$, but the WD confusion noise dominates the noise curve below $\sim 0.1 \,{\rm{Hz}}$. As we will quantitatively discuss later including astrophysical confusion noise, the BBO sensitivity in amplitude to a NS binary is nearly three times better than that of DECIGO.

\subsection{Astrophysical foregrounds}

DECIGO and BBO are sensitive to a large number of astrophysical GW sources, in particular NS, BH, and WD binaries. It is expected that the low-frequency side of the noise curve would be dominated by the GWs from cosmological (extra-galactic) population of WD binaries, which still remains after subtracting individually identified signals in the process of data analysis and are stochastic in nature. According to the estimation by \cite{bib7}, the fitting formula for the power spectrum of WD confusion noise \cite{bib12} (residual contribution after the subtraction) is given by
\begin{align}
S_h^{\rm{WD}}(f) &= 4.2 \times 10^{-47} \left( \frac{f}{1\,{\rm{Hz}}} \right)^{-7/3}\, \nonumber \\
& \times \exp \left[-2 \left( \frac{f}{5\times 10^{-2}\,{\rm{Hz}}} \right)^2 \right] \;\; {\rm{Hz}}^{-1}\,\;. \nonumber
\end{align}
There is also contribution from galactic WD binaries. However, as pointed out in \cite{bib14}, the galactic sources are sufficiently sparse in frequency space above $3\times 10^{-3} \,{\rm{Hz}}$ so that one is able to fit them out of the data. Actually, using the expression for the residual foreground after the subtraction given in \cite{bib12}, we verified that the galactic sources contribute less than 10 \% to the total-noise curve above the frequency $5\times 10^{-3}\,{\rm{Hz}}$. Thus, galactic contribution can be ignored for our present purpose.

Another astrophysical source that we have to take into account is the NS-binary foreground. According to \cite{bib14}, the energy density of GWs from NS binaries per logarithmic frequency bin normalized by the critical energy density of the universe at present is written as
\begin{align}
\Omega_{\rm{gw}}^{\rm{NS}} (f) &= \frac{8 \pi^{5/3}}{9 H_0^2} M_c^{5/3} f^{2/3} n_0 \;, \nonumber \\
n_0 &= \int_0^{\infty} \frac{\dot{n}(z)}{(1+z)^{4/3} H(z)} dz \;. 
\label{eq3}   
\end{align}
$H_0$ is the Hubble constant and $M_c$ is the chirp mass defined as $M_c \equiv \eta^{3/5} M_t$, together with the total mass $M_t=m_1+m_2$ and the symmetric 
mass ratio $\eta=m_1 m_2/M_t^2$. $\dot{n}(z)$ is the NS merger rate per unit comoving volume per unit proper time at a redshift $z$. We adopt the following fitting form of the NS-NS merger rate 
given in \cite{bib15}:  
\begin{align}
\dot{n}(z) &= \dot{n}_0\, s(z)\,\, ;\quad
s(z) = \left\{        
\begin{array}{ll}
1+2 z  & (z \leq 1) \\
\frac{3}{4} (5-z) &(1 < z \leq 5) \\
 0 & (5 < z)
\end{array}
\right. \;, 
\label{eq1}
\end{align}
where the function $s(z)$ is estimated based on the star formation history 
inferred from the UV luminosity \cite{bib16}. 
The quantity $\dot{n}_0$ represents the merger rate at present. We assume the flat $\Lambda$CDM universe with $\Omega_{\rm{m}}=0.3$ ($\Omega_{\Lambda}=0.7$) and the Hubble parameter $H(z)$ is given by 
\begin{equation}
H(z) = H_0 \left[ \Omega_m (1+z)^3 + (1-\Omega_m ) \right]^{1/2}. 
\label{eq2}
\end{equation}
Writing $H_0=h_{72} \times 72\,{\rm{km}}\,{\rm{s}}^{-1}\,{\rm{Mpc}}^{-1}$ and substituting Eqs.~(\ref{eq1}) and (\ref{eq2}) for Eq.~(\ref{eq3}), we obtain 
\begin{align}
\Omega_{\rm{gw}}^{\rm{NS}} (f) &= 3.74 \times 10^{-11} h_{72}^{-3} \nonumber \\
\times &\left( \frac{M_c}{1.22 M_{\odot}} \right)^{5/3} \left( \frac{f}{1\,{\rm{Hz}}} \right)^{2/3} \left( \frac{\dot{n}_0}{10^{-6}\,{\rm{Mpc}}^{-3}\,{\rm{yr}}^{-1}} \right) \;,  \nonumber
\end{align}
which is translated into the NS confusion-noise power spectrum 
\begin{align}
S_h^{\rm{NS}} (f) &= 1.55 \times 10^{-47} h_{72}^{-1} \nonumber \\
\times & \left( \frac{M_c}{1.22 M_{\odot}} \right)^{5/3} \left( \frac{f}{1\,{\rm{Hz}}} \right)^{-7/3} \left( \frac{\dot{n}_0}{10^{-6}\,{\rm{Mpc}}^{-3}\,{\rm{yr}}^{-1}} \right) \;, \nonumber 
\end{align}
using the relation
\begin{equation}
S_{\rm{h}}^{\rm{NS}}(f) = \frac{3H_0^2}{4\pi^2 f^3} \Omega_{\rm{gw}}^{\rm{NS}}(f) \;.
\label{eq16}
\end{equation}
BH-BH and NS-BH binaries are also the sources of the foreground. However, the populations are smaller, compared with NS-NS binaries. Therefore we neglect them for simplicity.

\subsection{Total noise curve}

DECIGO and BBO have almost the same sensitivities, but BBO is approximately three-times more sensitive to a GW from a NS binary than DECIGO in terms of signal-to-noise ratio (SNR) (e.g. for a NS binary at the redshift $z=5$, sky-position- and inclination-averaged SNR is $8.5$ and $26.3$ for a single interferometer of DECIGO and BBO, respectively.). In other words, the DECIGO noise curve improved by a factor 3 in amplitude is comparable to that of BBO. Our purpose in this paper is to assess cosmology achieved by space-based detectors such as DECIGO and BBO, and to make the requirement for the experimental design clear. Thus, we consider only DECIGO, but allowing the noise curve in Eq.~(\ref{eq4}) to vary by an overall factor $r_n=1$, $1/2$, $1/3$, $1/5$ in amplitude. Namely, the power spectrum scales as
\begin{equation}
S_h^{\rm{inst}} (f) = r_n^2 \times \hat{S}_{\rm{h,\,D}}^{\rm{inst}} (f) \,. \nonumber
\end{equation}
Note again that BBO noise power spectrum in Eq.~(\ref{eq5}) approximately corresponds to that of DECIGO with $r_n=1/3$. 

\begin{figure}[t]
\begin{center}
\includegraphics[width=8.5cm]{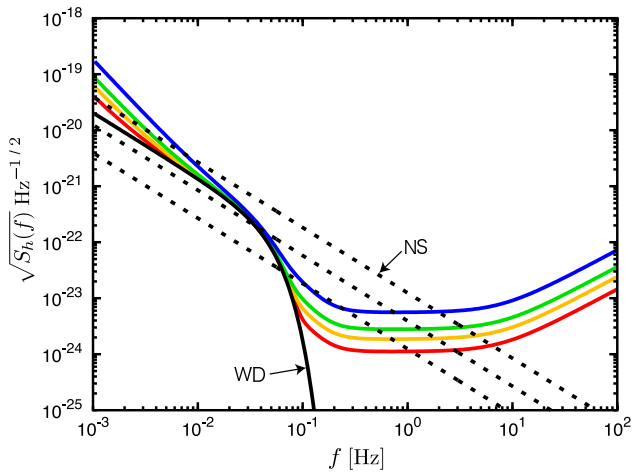}
\caption{Noise curves of default DECIGO (solid blue curve) and those scaled by $r_n=1/2$ (green), $1/3$ (orange), $1/5$ (red), which include the confusion noise from a number of WD binaries. The WD-binary foreground $\sqrt{S_h^{\rm{WD}}}$ is shown with solid, black curve. The three diagonal lines represent the NS-binary foreground $\sqrt{S_h^{\rm{NS}}}$ before subtraction (or ${\cal{R}}_{\rm{NS}}=1$) with $\dot{n}_0=10^{-5},\,10^{-6},\,10^{-7}\,{\rm{Mpc}}^{-3}\,{\rm{yr}}^{-1}$ from the top to the bottom.}
\label{fig2}
\end{center}
\end{figure}

Including astrophysical contributions, a total-noise power spectrum is given by 
\begin{equation}
S_h(f)=S_h^{\rm{inst}}(f) + S_h^{\rm{WD}}(f)+ S_h^{\rm{NS}}(f) {\cal{R}}_{\rm{NS}} \;. 
\label{eq6}
\end{equation}
Here the factor ${\cal{R}}_{\rm{NS}}$ denotes a suppression factor due to the subtraction of individually identified NS binaries, which will be estimated in Sec.~\ref{sec4}. The contribution of each term in Eq.~(\ref{eq6}) is shown in Fig.~\ref{fig2}. At high frequencies, roughly above $0.2\,{\rm{Hz}}$, the noise curves scale straightforward with $r_n$ if we neglect the contribution from NS binaries. On the other hand, at low frequencies, the WD confusion noise prevents the total noise from improving as $r_n$ decreases. This means that, from the experimental point of view, improvement required for the instrument is involved with not radiation-pressure noise but laser shot noise or beam-pointing-jitter and stray light noises.

\subsection{Outline of analysis and noise model}
In this paper, we deal with the instrumental noise and the astrophysical confusion noise in a self-consistent manner when we estimate the sensitivities to dark energy and primordial GW backgrounds. Firstly, in Sec.~\ref{sec4}, we compute SNR for each NS binary with the noise power spectrum, $S_h^{\rm{inst}}(f) + S_h^{\rm{WD}}(f)$, assuming that large signals well above the noise curve are easily subtracted. Then, we determine how much the residual amplitude of the NS-binary foreground remains after the subtraction process, which depends on the SNR threshold value. In the next step, the noise spectrum, $S_h^{\rm{inst}}(f) + S_h^{\rm{WD}}(f)+ S_h^{\rm{NS}}(f) {\cal{R}}_{\rm{NS}}$, is used to calculate the Fisher matrices for parameter estimation of NS binaries and cosmological models. The contribution of the confusion noise, $S_h^{\rm{WD}}(f)+ S_h^{\rm{NS}}(f) {\cal{R}}_{\rm{NS}}$, is also taken into account in the correlation analysis for a stochastic GW background.

\section{GW waveform of a binary and SNR}
\label{sec3}
For a single binary system, the Fourier transform of the GW waveform is 
expressed as a function of frequency $f$ \cite{bib18,bib19},
\begin{equation}
\tilde{h} (f) = \frac{A}{d_L(z)} M_z^{5/6} f^{-7/6} e^{i \Psi(f)} \;, 
\label{eq9}
\end{equation}
where $d_L$ is the luminosity distance, and  
the quantity $M_z=(1+z) M_c$ is the 
redshifted chirp mass. $M_c$ is the proper chirp mass defined in the source rest frame. The constant $A$ is given by $A= (\sqrt{6}\, \pi^{2/3})^{-1}$, 
which includes the factor $\sqrt{4/5}$ for a geometrical 
average over the inclination angle of a binary \cite{bib80}. The function $\Psi(f)$ represents the frequency-dependent phase arising from 
the orbital evolution, and at the order of the {\it{restricted}} $1.5$ 
post-Newtonian (PN) approximation, it is given by \cite{bib18,bib19} 
\begin{align}
\Psi (f) &= 2\pi f \,t_c -\phi_c -\frac{\pi}{4} +\frac{3}{128} 
(\pi M_z f)^{-5/3} \nonumber \\
& \times \left[ 1+ \frac{20}{9} \left(\frac{743}{336} + 
\frac{11}{4} \eta \right)
\eta^{-2/5} (\pi M_z f)^{2/3} \right. \nonumber \\
& \left. - 16 \pi \eta^{-3/5} (\pi M_z f) - \frac{25}{768} X(z) M_z (\pi f M_z)^{-8/3} \right] \;, 
\label{eq8}
\end{align}
where $\eta=m_1 m_2/(m_1+m_2)^2$ is the symmetric mass ratio, and $t_c$ and $\phi_c$ are the time and phase at coalescence, 
respectively. The first term in the square brackets in Eq.~(\ref{eq8}) 
corresponds to the Newtonian-order dynamics 
and the second and third terms represent the post Newtonian-order corrections in powers of 
$v \sim (\pi M_z f)^{1/3}$. The last term in the bracket is a phase correction due to cosmic expansion \cite{bib1,bib20}, where $X(z)$ is defined as
\begin{equation}
X(z) \equiv \frac{1}{2} \left( H_0 -\frac{H(z)}{1+z} \right) \;, \nonumber
\end{equation}
or equivalently expressed as $[ \dot{a}(0)-\dot{a}(z) ]/2$. Thus, $X(z)>0$ ($X(z)<0$) corresponds to the accelerating (decelerating) universe. Note that this correction is "$-4$ PN" order and becomes more important at lower frequencies. This is because a binary longer stays at low frequencies (when the binary separation is larger) and has the larger cycle number. 

The squared SNR of a binary GW signal at a redshift $z$ is defined by
\begin{equation}
\bar{\rho}^2(z) \equiv 4 \sum_{i=1}^8 \int_{f_{\rm{min}}}^{f_{\rm{max}}} \frac{|\tilde{h}(f)|^2}{S_h(f)} df \;.
\label{eq7}
\end{equation}
Here $S_{\rm{h}}(f)$ and $\tilde{h} (f)$ are given by Eqs.~(\ref{eq6}) and (\ref{eq9}), respectively. The summation in Eq.~(\ref{eq7}) is taken with respect to the number of independent detectors of DECIGO. (DECIGO has four clusters, each of which has two independent interferometers). Since it is assumed that all interferometers have identical noise-levels, the summation can be replaced just with an overall factor of 8. The lower cutoff frequency $f_{\rm min}$ is given by the function of observation 
time $T_{\rm obs}$ as well as the redshift and mass: 
\begin{equation}
f_{\rm{min}} = 0.233 \left( \frac{1M_{\odot}}{M_z} \right)^{5/8} 
\left( \frac{1\,{\rm{yr}}}{T_{\rm{obs}}} \right)^{3/8} \; {\rm{Hz}}\;. 
\label{eq10}
\end{equation}
The upper cutoff of the frequency $f_{\rm{max}}$ naturally arises from the noise curve, since the coalescence frequency of the binary with the mass $\sim M_{\odot}$ is 
typically $\sim {\rm{kHz}}$. For the computational 
purpose, we set $f_{\rm{max}}=100$\,Hz.

\section{Neutron-star binary subtraction}
\label{sec4}

NS-binary foreground is not stochastic in the DECIGO frequency band \cite{bib15} and in principle can be subtracted from observation data. How well one can subtract them and how much the residuals are left depend on the estimation accuracy of binary parameters, or the SNR of the signal from each NS binary. This subtraction process is mandatory to use NS binaries as a probe for dark energy and to search for a primordial GW background.

In a recent paper \cite{bib17}, Yagi and Seto have performed Monte-Carlo simulations generating synthetic GW data for binaries at random sky positions and with orientations of angular-momentum axes over redshifts up to $z=5$. They estimated how much residuals are left after the subtraction process. The criterion for the successful subtraction is that SNR $\rho_i$ for each binary exceeds the SNR threshold $\rho_{\rm{th}}$. The fraction of the residuals in the energy density of NS binaries after the subtraction can be expressed as
\begin{align}
{\cal{R}}_{\rm{NS}} &= \frac{\int_0^{\infty} dz \, F(z) \Delta \Omega_{\rm{gw}}^{\rm{NS}} (f,z)}{\int_0^{\infty} dz \,\Delta \Omega_{\rm{gw}}^{\rm{NS}} (f,z)}\;, \nonumber \\
F(z) &\equiv \frac{\sum_i \rho_i^2 (z) \Theta [ \rho_{\rm{th}}-\rho_i (z) ]}{\sum_i \rho_i^2 (z)} \;, \nonumber
\end{align}
together with 
\begin{equation}
\Delta \Omega_{\rm{gw}}^{\rm{NS}} (f,z) = \frac{8 \pi^{5/3}}{9 H_0^2} M_c^{5/3} f^{2/3} \frac{\dot{n}(z)}{(1+z)^{4/3} H(z)} \;, \nonumber 
\end{equation} 
where $\Theta [\cdots]$ is the step function. 

\begin{figure}[t]
\begin{center}
\includegraphics[width=8cm]{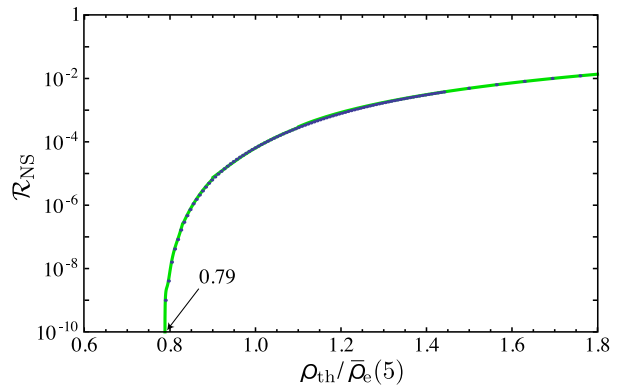}
\caption{Fraction of NS residual ${\cal{R}}_{\rm{NS}}$ after the subtraction as a function of $\rho_{\rm{th}}/\bar{\rho}_e(5)$. The fitting formula is presented in Appendix \ref{app1}.}
\label{fig3}
\end{center}
\end{figure}

In Ref. \cite{bib17}, the fractional residual ${\cal{R}}_{\rm{NS}}$ is provided as a function of $\rho_{\rm{th}}/\bar{\rho}_{\rm{e}}(5)$, where $\bar{\rho}_{\rm{e}}(5)$ is the SNR for an edge-on NS binary at the redshift $z=5$. The ${\cal{R}}_{\rm{NS}}$ is shown in Fig.~\ref{fig3}. The figure indicates that ${\cal{R}}_{\rm{NS}}$ rapidly decreases at the point where $\rho_{\rm{th}}/\bar{\rho}_{\rm{e}}(5) \approx 0.79$ or equivalently $\bar{\rho}_{\rm{e}}(5)$ is nearly 30\% better than $\rho_{\rm{th}}$. In Appendix \ref{app1}, based on this result, we give an explicit fitting form of the result for the later calculation of ${\cal{R}}_{\rm{NS}}$. 

The $\bar{\rho}_{\rm{e}}(5)$ is related with the inclination-angle averaged SNR $\bar{\rho}(5)$ defined in Eq.~(\ref{eq7}) by $\bar{\rho}_{\rm{e}}(5)=\sqrt{5}/4\times \bar{\rho}(5)$. This is because the angular dependences of GW waveforms are $h_+ \propto (1+\cos^2 \theta)/2$ and $h_{\times} \propto \cos \theta$ for each polarization mode, where $\theta$ is the angle between the line of sight and the axis of the orbital motion of the binary. In order to calculate $\bar{\rho}(5)$, we use the noise power spectrum in Eq.~(\ref{eq6}) without the NS-binary contribution. Strictly speaking, the NS-binary foreground should be initially included and each binary signal has to be iteratively subtracted one by one for the consistency of the calculation. However, as the authors of \cite{bib17} have performed, we can simplify the procedure by neglecting the contribution from the NS-binary foreground in the calculation of SNR. This simplification would be valid because large GW signals can be easily subtracted from the data to the noise level that is limited by the instrumental noise and the WD foreground. We choose the SNR threshold as $\rho_{\rm{th}}=20$.

Once $\bar{\rho}(5)$ is obtained, we can calculate $\rho_{\rm{th}}/\bar{\rho}_{\rm{e}}(5)$ and translate it into ${\cal{R}}_{\rm{NS}}$ by using the fitting formula in Appendix \ref{app1}. The values of ${\cal{R}}_{\rm{NS}}$ for observation time, $T_{\rm{obs}}=3,\, 5,\, 10\,{\rm{yr}}$, and noise-curve scaling, $r_n=1,\,1/2,\,1/3,\,1/5$, are listed in Table \ref{tab1}. If $r_n$ is less than a half, all NS binaries are fitted out. Even in the case of $r_n=1$, the amplitude of the residual of the NS foreground is $\sim \sqrt{200}\sim14$ times smaller than the initial foreground levels and hardly affect the shape of the noise curves presented in Fig.~\ref{fig2}. Note, however, that the suppression of the NS foreground below the noise curve is not always sufficient for the detection of a GW background, because the residual may contaminate a correlation signal. This issue is discussed in Sec.~\ref{sec6}.

\begin{table*}[t]
\begin{center}
\begin{tabular}{|c|c|c|c|c|c|c|c|c|c|}
\hline
& \multicolumn{3}{|c|}{$\quad T_{\rm{obs}}=3\,{\rm{yr}}\; (f_{\rm{min}}=0.0446\,{\rm{Hz}}) \quad $} & \multicolumn{3}{|c|}{$\quad T_{\rm{obs}}=5\,{\rm{yr}}\; (f_{\rm{min}}=0.0368\,{\rm{Hz}}) \quad$} & \multicolumn{3}{|c|}{$\quad T_{\rm{obs}}=10\,{\rm{yr}}\; (f_{\rm{min}}=0.0284\,{\rm{Hz}}) \quad$} \\
 \hline 
$r_n$ & $\quad \bar{\rho}(5) \quad$ & $\;\; \rho_{\rm{th}}/\bar{\rho}_e(5) \;\;$ & ${\cal{R}}_{\rm{NS}}$ & $\quad \bar{\rho}(5) \quad$ & $\;\; \rho_{\rm{th}}/\bar{\rho}_e(5) \;\;$ & ${\cal{R}}_{\rm{NS}}$ & $\quad \bar{\rho}(5) \quad$ & $\;\; \rho_{\rm{th}}/\bar{\rho}_e(5) \;\;$ & ${\cal{R}}_{\rm{NS}}$ \\ 
\hline
$1$ & $24.06$ & $1.487$ & $4.66\times 10^{-3}$ & $24.07$ & $1.486$ & $4.64 \times 10^{-3}$ & $24.09$ & $1.485$ & $4.62\times10^{-3}$ \\ 
$1/2$ & $47.75$ & $0.749$ & $0$ & $47.77$ & $0.749$ & $0$ & $47.78$ & $0.749$ & $0$ \\ 
$1/3$ & $71.25$ & $0.502$ & $0$ & $71.25$ & $0.502$ & $0$ & $71.26$ & $0.502$ & $0$ \\ 
$1/5$ & $118.9$ & $0.304$ & $0$ & $117.9$ & $0.303$ & $0$ & $117.9$ & $0.303$ & $0$ \\ 
\hline
\end{tabular}
\end{center}
\caption{Fraction of NS residual ${\cal{R}}_{\rm{NS}}$ for each parameter set of $T_{\rm{obs}}$ and $r_n$.}
\label{tab1}
\end{table*}

\section{Standard siren as a probe for dark energy}
\label{sec5}

\subsection{Standard siren}

The continuous GW signal from a compact-binary object provides a unique way to measure the luminosity 
distance to the source with high precision \cite{bib6}. Such binary sources are often 
referred to as standard sirens. There are two different aspects in using the standard siren. 

One is to measure the amplitude of GWs, or the luminosity distance as a function of redshifts (see Eq.~(\ref{eq9})). However, GW information alone cannot provide a source redshift, since the mass parameter we can determine is $M_z=(1+z)M_c$ and the source redshift degenerates with the proper source mass $M_c$ \cite{bib91}. So the redshift has to be determined from electromagnetic observation of the host galaxy. This method as a tool to measure the cosmic expansion has been investigated by many authors \cite{bib4,bib21,bib22,bib23,bib24,bib25,bib26,bib27,bib28}, assuming that source redshifts are known by spectroscopic follow-up observations. Thus, the feasibility of using NS binaries as the standard siren relies on the determination of the redshift of each binary, which requires high-angular resolution for GW detectors to select a true host galaxy out of several candidates. A lack of angular resolution of ground-based detectors \cite{bib77,bib78,bib79} or Laser Interferometer Space Antenna (LISA) \cite{bib23,bib27} limits the availability of the standard sirens as a cosmological probe, though the identification of the host galaxy is possible for the binary event accompanying an electromagnetic counterpart. On the other hand, according to Cutler and Holz \cite{bib4}, BBO has the angular resolution $\sim 1-100\,{\rm{arcsec}}^2$ and can uniquely identify the host galaxy of the binary. The authors also have shown that cosmological parameters can be measured with high precision, using a large number of NS binaries, $\sim 10^6$.

Another method utilizes the phase modulation of GWs due to the cosmic expansion, which is the last term including $X(z)$ in Eq.~(\ref{eq8}) \cite{bib1,bib20}. Since the amplitude and phase bring us independent information, this method is complementary to the method described above. While the redshifts of the binaries are assumed to be determined by the electromagnetic follow-ups in the preceding works \cite{bib1,bib20}, we point out in this paper for the first time that we can measure a cosmic-expansion rate only with GW observations without electromagnetically-estimated redshifts by combining the luminosity distance and the phase shift. This is a great advantage of our method, because a pure GW observation enables us to compare its observational data with those obtained in other electromagnetic observations. However, unfortunately, this method is not applicable to advanced ground-based detectors \cite{bib58,bib59,bib60,bib61} and LISA as a sensitive cosmological probe, because of the weakness of the signal coming from the cosmic expansion. To detect the signal, a large number of GW sources is necessary. DECIGO and BBO are expected to detect a lot of NS binaries, which enable them to detect the weak signal. In the subsequent sections, we first investigate the power of utilizing the phase correction, without assuming any electromagnetically-estimated redshift, and give the figure of merits. Then we will compare this method with the previous one, in which source redshifts are assumed to be given, and combine both methods to maximize the accuracy of the estimation.

\subsection{Constraints on cosmological parameters}
\label{sec5b}

\begin{figure*}[t]
\begin{center}
\includegraphics[width=16cm]{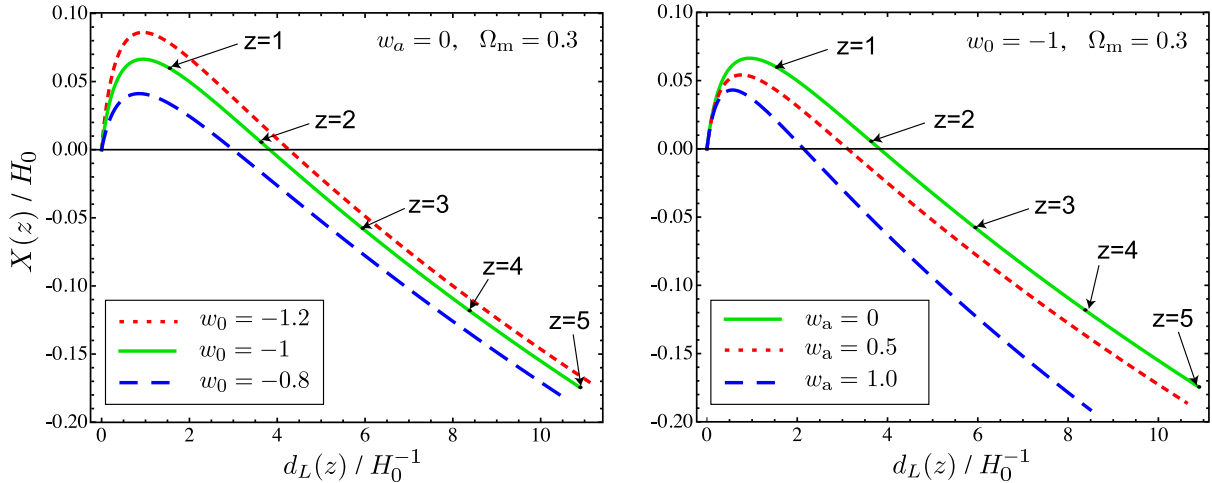}
\caption{$d_L$ - $X$ plot as a function of redshift. Left: $w_0$ is varied, while $w_a$ and $\Omega_{\rm{m}}$ are fixed. Each curve corresponds to $w_0=-1.2$ (red, dotted curve), $w_0=-1$ (green, solid curve), $w_0=-0.8$ (blue, dashed curve). Right: $w_a$ is varied, while $w_0$ and $\Omega_{\rm{m}}$ are fixed. Each curve corresponds to $w_a=0$ (green, solid curve), $w_a=0.5$ (red, dotted curve), $w_a=1.0$ (blue, dashed curve). The green solid curves in both figures are for $\Lambda$CDM universe. The points corresponding to $z=1,2,3,4,5$ are shown on the curve.}
\label{fig4}
\end{center}
\end{figure*}

For simplicity, we consider a spatially flat universe with dark energy whose equation of state is parametrized as $w(z) = w_0 +w_a z/(1+z)$ \cite{bib93}. Then the luminosity distance and the cosmological phase correction in Eqs.~(\ref{eq9}) and (\ref{eq8}) are written as
\begin{align}
d_L(z) &= (1+z) \int_{0}^{z} \frac{dz^{\prime}}{H(z^{\prime})} \;, 
\label{eq17} \\
X(z) &= \frac{1}{2} \left( H_0 -\frac{H(z)}{1+z} \right) \;, 
\label{eq11} \\
H(z) &= H_0 \left\{ \Omega_m (1+z)^3 \right. \nonumber \\
& \left. + (1-\Omega_m ) (1+z)^{3(1+w_0+w_a)} \exp \left[ - \frac{3 w_a z}{1+z} \right] \right\}^{1/2} \;. \nonumber
\end{align}
In the above, $w_0$, $w_a$, $\Omega_m$, and $H_0$ are the parameters to be determined from GW observations. For the analysis of error estimation, we adopt a fiducial set of cosmological parameters: $w_0=-1$, $w_a=0$, $\Omega_m=0.3$, $H_0=72\,{\rm{km}}\,{\rm{s}}^{-1}\,{\rm{Mpc}}^{-1}$. For illustrative purpose, we give some $d_L$ - $X$ plots for certain parameter sets in Fig.~\ref{fig4}. The redshift dependence of the curves are different depending the values of $w_0$ and $w_a$ so that in principle one can independently determine these parameters.    

In what follows, we calculate the estimation errors of binary parameters $\theta _a$: $M_z$, $\eta$, $t_c$, $\phi_c$, $d_L$, and $X$ in the waveform of Eq.~(\ref{eq9}). Then we derive the estimation errors of the cosmological parameters from them. The fundamental tool we use to estimate the errors for a single binary 
is the Fisher matrix formalism. The Fisher matrix for a single binary 
is given by \cite{bib18,bib29}
\begin{align}
\Gamma_{ab}^{({\rm{single}})} &= 4 \sum_{i=1}^{8}\, {\rm{Re}} \int_{f_{\rm{min}}}^{f_{\rm{max}}}
 \frac{\partial_{a} \tilde{h}^{\ast}(f)\, \partial_{b}
 \tilde{h}(f)}{S_{\rm{h}}(f)} df \;,
 \label{eq13}
\end{align}
where $\partial_a$ denotes a derivative with respect to a parameter $\theta_a$. Since the eight interferometers of DECIGO and BBO are identical, the summation is reduced to just multiplying by a factor of 8. The noise power spectrum and GW signal are given by Eqs.~(\ref{eq6}) and (\ref{eq9}), respectively. The frequency cutoffs are also given by Eq.~(\ref{eq10}) and $f_{\rm{max}}=100\,{\rm{Hz}}$. Given the numerically evaluated Fisher matrix, 
the marginalized 1-$\sigma$ error of a parameter, $\Delta\theta_a$,
is estimated from the inverse Fisher matrix 
\begin{align}
\Delta \theta_a = \sqrt{\{\mathbf{\Gamma}^{-1}\}_{aa}}. \nonumber
\end{align}

Once the estimation error of $X$ is obtained, it is straightforward to calculate the estimation errors of the cosmological parameters. As described in Appendix \ref{app2}, the error of $d_L$ is much smaller than that of $X$. This fact means that we can replace an observed $d_L$ with the corresponding redshift in the fiducial cosmological model when we derive the measurement accuracies of cosmological parameters from $X$. Thus, for the simplicity of the analysis, we use $X(z)$ instead of $X(d_L)$. Futhermore, we assume that the Hubble constant $H_0$ is known {\it{a priori}}, and fix it to $H_0=72\,{\rm{km}}\,{\rm{s}}^{-1}\,{\rm{Mpc}}^{-1}$, because the Hubble constant has been determined at a-few-percent level from the observation of nearby Cepheids and supernovae \cite{bib32}. Thus, the free parameters of the Fisher matrix are $w_0$, $w_a$, and $\Omega_m$. The Fisher matrix is given by
\begin{equation}
\Gamma_{ab} = \int_{0}^{\infty} \frac{\partial_a X(z) \partial_b X(z)}{\sigma_X^2(z)} \frac{d N(z)}{dz} dz\;, 
\label{eq12}
\end{equation}
where $dN(z)/dz$ is the number of NS binaries in the redshift interval $[z,z+dz]$ observed during $T_{\rm obs}$ and is given by \cite{bib15}
\begin{equation}
\frac{dN(z)}{dz}=T_{\rm obs}\,\frac{4\pi r^2(z)}{H(z)}\frac{\dot{n}(z)}{1+z} \;, 
\label{eq20}
\end{equation}
where $r(z)$ is the comoving radial distance defined as $r(z)=d_L(z)/(1+z)$ and $\dot{n}$ is given in Eq.~(\ref{eq1}). Since the normalization of $\dot{n}$ is still uncertain, we adopt the most recent estimate, 
$\dot{n}_0=10^{-6}\,{\rm{Mpc}}^{-3}\, {\rm{yr}}^{-1}$, as 
a reliable estimate based on extrapolations from the 
observed binary pulsars in our Galaxy \cite{bib30}, and also consider optimistic and pessimistic values, $10^{-5},\,10^{-7}\,{\rm{Mpc}}^{-3}\, {\rm{yr}}^{-1}$. Since the observation time $T_{\rm{obs}}$ is also a crucial parameter, we suppose 3-yr, 5-yr, and 10-yr observation for the Fisher matrix. In Eqs.~(\ref{eq12}) and (\ref{eq20}), the integration is performed explicitly with respect to a redshift. Note, however, that the integrand can be regarded as a function of the luminosity distances, instead of the redshifts, as discussed in Appendix \ref{app2}. 

\begin{table*}[t]
\begin{center}
\begin{tabular}{|c|c|c|c|c|c|c|c|c|c|}
\hline
\multicolumn{10}{|c|}{$T_{\rm{obs}}=3\,{\rm{yr}}$} \\
\hline
& \multicolumn{3}{|c|}{$\quad \quad \quad \quad \dot{n}_0=10^{-7} \quad \quad \quad \quad$} & \multicolumn{3}{|c|}{$\quad \quad \quad \quad \dot{n}_0=10^{-6} \quad \quad \quad \quad$} & \multicolumn{3}{|c|}{$\quad \quad \quad \quad \dot{n}_0=10^{-5} \quad \quad \quad \quad$} \\
\hline 
$r_n$ & $\quad \Delta w_0 \quad$ & $\quad \Delta w_a \quad$ & FoM & $\quad \Delta w_0 \quad$ & $\quad \Delta w_a \quad$ & FoM & $\quad \Delta w_0 \quad$ & $\quad \Delta w_a \quad$ & FoM \\
\hline
$1$ & $3.05 \times 10^{0}$ & $2.63 \times 10^{1}$ & $1.71\times10^{-2}$ & $9.81\times 10^{-1}$ & $8.44\times 10^{0}$ & $1.66\times10^{-1}$ & $3.59\times 10^{-1}$ & $2.99\times 10^{0}$ & $1.33\times10^{0}$ \\ 
$1/2$ & $1.53\times 10^{0}$ & $1.37\times 10^{1}$ & $6.32\times10^{-2}$ & $5.21\times 10^{-1}$ & $4.59\times 10^{0}$ & $5.65\times10^{-1}$ & $1.53\times 10^{-1}$ & $1.37\times 10^{0}$ & $6.32\times10^{0}$ \\ 
$1/3$ & $1.03\times 10^{0}$ & $9.55\times 10^{0}$ & $1.31\times10^{-1}$ & $3.25\times 10^{-1}$ & $3.02\times 10^{0}$ & $1.31\times10^{0}$ & $1.03\times 10^{-1}$ & $9.56\times 10^{-1}$ & $1.31\times10^{1}$ \\ 
$1/5$ & $6.26\times 10^{-1}$ & $6.15\times 10^{0}$ & $3.20\times10^{-1}$ & $1.98\times 10^{-1}$ & $1.95\times 10^{0}$ & $3.20\times10^{0}$ & $6.26\times10^{-2}$ & $6.15\times 10^{-1}$ & $3.20\times10^{1}$ \\ 
\hline
\multicolumn{10}{c}{} \\
\hline
\multicolumn{10}{|c|}{$T_{\rm{obs}}=5\,{\rm{yr}}$} \\
\hline
& \multicolumn{3}{|c|}{$\dot{n}_0=10^{-7}$} & \multicolumn{3}{|c|}{$\dot{n}_0=10^{-6}$} & \multicolumn{3}{|c|}{$\dot{n}_0=10^{-5}$} \\
\hline 
$r_n$ & $\Delta w_0$ & $\Delta w_a$ & FoM & $\Delta w_0$ & $\Delta w_a$ & FoM & $\Delta w_0$ & $\Delta w_a$ & FoM \\
\hline
$1$ & $1.03\times 10^{0}$ & $1.02\times 10^{1}$ & $1.18\times10^{-1}$ & $3.30\times 10^{-1}$ & $3.25\times 10^{0}$ & $1.14\times10^{0}$ & $1.20\times 10^{-1}$ & $1.14\times 10^{0}$ & $9.21\times10^{0}$ \\ 
$1/2$ & $5.24\times 10^{-1}$ & $5.46\times 10^{0}$ & $4.10\times10^{-1}$ & $1.66\times 10^{-1}$ & $1.73\times 10^{0}$ & $4.10\times10^{0}$ & $5.24\times10^{-2}$ & $5.46\times 10^{-1}$ & $4.10\times10^{1}$ \\ 
$1/3$ & $3.59\times 10^{-1}$ & $3.89\times 10^{0}$ & $8.12\times10^{-1}$ & $1.13\times 10^{-1}$ & $1.23\times 10^{0}$ & $8.13\times10^{0}$ & $3.59\times10^{-2}$ & $3.89\times 10^{-1}$ & $8.12\times10^{1}$ \\ 
$1/5$ & $2.28\times 10^{-1}$ & $2.61\times 10^{0}$ & $1.82\times10^{0}$ & $7.20\times10^{-2}$ & $8.26\times 10^{-1}$ & $1.82\times10^{1}$ & $2.28\times10^{-2}$ & $2.61\times 10^{-1}$ & $1.82\times10^{2}$ \\ 
\hline 
\multicolumn{10}{c}{} \\
\hline
\multicolumn{10}{|c|}{$T_{\rm{obs}}=10\,{\rm{yr}}$} \\
\hline
& \multicolumn{3}{|c|}{$\dot{n}_0=10^{-7}$} & \multicolumn{3}{|c|}{$\dot{n}_0=10^{-6}$} & \multicolumn{3}{|c|}{$\dot{n}_0=10^{-5}$} \\
\hline 
$r_n$ & $\Delta w_0$ & $\Delta w_a$ & FoM & $\Delta w_0$ & $\Delta w_a$ & FoM & $\Delta w_0$ & $\Delta w_a$ & FoM \\
\hline
$1$ & $2.57\times 10^{-1}$ & $2.84\times 10^{0}$ & $1.56\times10^{0}$ & $8.25\times10^{-2}$ & $9.06\times 10^{-1}$ & $1.53\times10^{1}$ & $2.94\times10^{-2}$ & $3.15\times 10^{-1}$ & $1.26\times10^{2}$ \\ 
$1/2$ & $1.40\times 10^{-1}$ & $1.64\times 10^{0}$ & $4.73\times10^{0}$ & $4.43\times10^{-2}$ & $5.17\times 10^{-1}$ & $4.72\times10^{1}$ & $1.40\times10^{-2}$ & $1.64\times 10^{-1}$ & $4.73\times10^{2}$ \\ 
$1/3$ & $1.03\times 10^{-1}$ & $1.23\times 10^{0}$ & $8.33\times10^{0}$ & $3.24\times10^{-2}$ & $3.90\times 10^{-1}$ & $8.33\times10^{1}$ & $1.03\times10^{-2}$ & $1.23\times 10^{-1}$ & $8.33\times10^{2}$ \\ 
$1/5$ & $7.28\times10^{-2}$ & $8.99\times 10^{-1}$ & $1.57\times10^{1}$ & $2.30\times10^{-2}$ & $2.84\times 10^{-1}$ & $1.57\times10^{2}$ & $7.28\times10^{-3}$ & $8.99\times10^{-2}$ & $1.57\times10^{3}$ \\ 
\hline
\end{tabular}
\end{center}
\caption{Measurement accuracy of $w_0$ and $w_a$, and FoM defined in Eq.~(\ref{eq30}) for each parameter set of $T_{\rm{obs}}$, $\dot{n}_0$, and $r_n$. In the table, $\dot{n}_0$ is in the unit of ${\rm{Mpc}}^{-3}\,{\rm{yr}}^{-1}$.}
\label{tab3}
\end{table*}

\subsection{Results}

\begin{figure*}[t]
\begin{center}
\includegraphics[width=15.5cm]{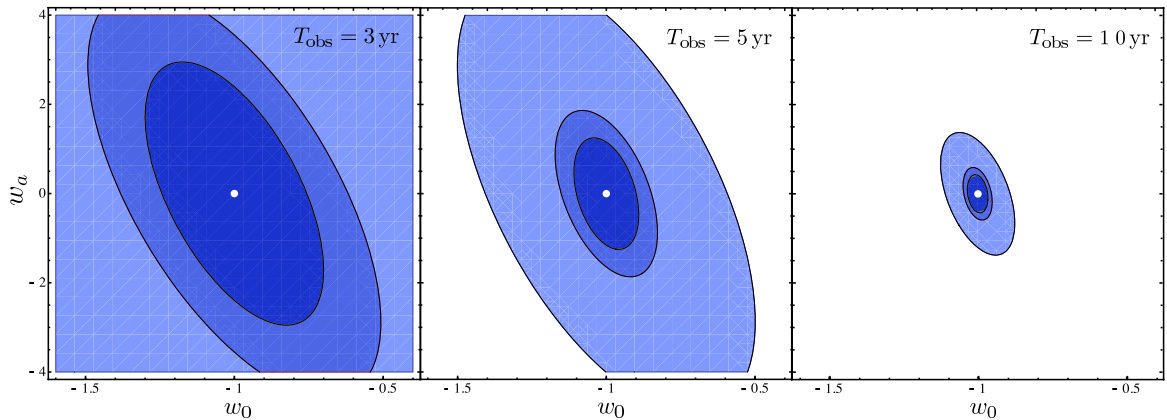}
\caption{$w_0$ - $w_a$ error ellipses marginalized over $\Omega_m$. The merger rate is fixed to $\dot{n}_0=10^{-6}\,{\rm{Mpc}}^{-3}\,{\rm{yr}}^{-1}$. From the left to the right panels, the observation time is $3\,{\rm{yr}}$, $5\,{\rm{yr}}$, and $10\,{\rm{yr}}$. In each panel, the larger to smaller ellipses denote those with $r_n=1$, $1/3$, and $1/5$, respectively. The dot at the center is our fiducial value: $w_0=-1$ and $w_a=0$.}
\label{fig6}
\end{center}
\end{figure*}

Since we are interested in parameters involved with dark energy, the measurement accuracies of $w_0$ and $w_a$ marginalized over the other remaining parameter are tabulated in Table \ref{tab3}. They change significantly depending on the parameters, $\dot{n}_0$, $T_{\rm{obs}}$, and $r_n$. For example, if we choose the intermediate values, $\dot{n}_0=10^{-6}\,{\rm{Mpc}}^{-3}\,{\rm{yr}}^{-1}$ and $T_{\rm{obs}}=5\,{\rm{yr}}$, the errors $\Delta w_0$ and $\Delta w_a$ are 0.330 and 3.254 for $r_n=1$ and 0.113 and 1.231 for $r_n=1/3$. For 10-yr observation, $\Delta w_0$ and $\Delta w_a$ reach at the levels of several $\times 10^{-2}$ and a few $\times 10^{-1}$, respectively. Error ellipses on the $w_0$ - $w_a$ plane are shown in Fig.~\ref{fig6} for the fixed $\dot{n}_0$  varying the values of $r_n$ and $T_{\rm{obs}}$. 

\begin{figure*}[t]
\begin{center}
\includegraphics[width=16.5cm]{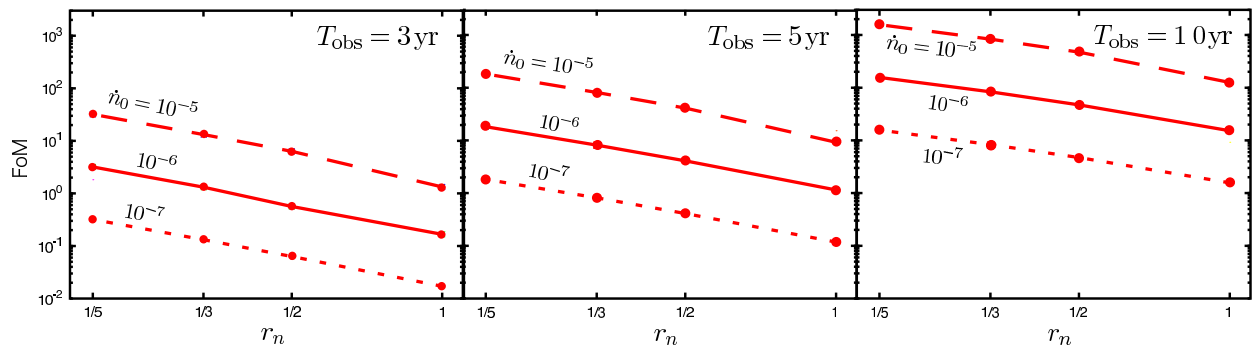}
\caption{FoM achievable for each parameter set of $r_n$, $T_{\rm{obs}}$, and $\dot{n}_0$. From the left to the right panels, the observation time is $3\,{\rm{yr}}$, $5\,{\rm{yr}}$, and $10\,{\rm{yr}}$. In each panel, the merger rate is represented by curve types: $\dot{n}_0=10^{-5}$ (dashed), $10^{-6}$ (solid), $10^{-7}$ (dotted), in the unit of ${\rm{Mpc}}^{-3}\,{\rm{yr}}^{-1}$.}
\label{fig5}
\end{center}
\end{figure*}

Scaling relations of the measurement accuracy with $\dot{n}_0$, $T_{\rm{obs}}$, and $r_n$ help us understand the results. $\Delta w_0$ and $\Delta w_a$ are expected to scale like
\begin{equation}
\Delta w_0,\; \Delta w_a \propto \sqrt{\Gamma^{-1}} \propto \dot{n}_0^{-1/2} T_{\rm{obs}}^{-31/16} r_n \;. 
\label{eq21}
\end{equation}
This relation can be understood as follows. The $\dot{n}_0$ dependence purely comes from the number of NS binaries through $\dot{n}(z)$ in Eq.~(\ref{eq20}). The $r_n$ dependence is inferred from the noise spectrum $S_h$ in Eq.~(\ref{eq13}), which is proportional to $r_n^2$. The observational-time dependence is rather complicated and comes from two factors: the number of NS binaries and the GW phase correction involved with $X$. From Eq.~(\ref{eq12}), $\Gamma_{ab} \propto dN/dz \propto T_{\rm{obs}}$. On the other hand, from Eq.~(\ref{eq8}), the cosmological-phase-correction term is larger at low frequencies and is roughly proportional to $f_{\rm{min}}^{-13/3}$. Around the frequency $f_{\rm{min}}$, noise is dominated by the WD foreground, which is $S_h^{WD} \propto f^{-7/3}$ and effectively reduces the power of $f_{\rm{min}}$ in the phase-correction term. So, from Eq.~(\ref{eq13}), 
\begin{align}
\sigma_X^{-2} &= \Gamma_{XX}^{(\rm{single})} \propto \int_{f_{\rm{min}}} \frac{(f^{-7/6} f^{-13/3})^2}{f^{-7/3}} df \nonumber \\
&\sim f_{\rm{min}}^{-23/3} \propto T_{\rm{obs}}^{23/8} \;.
\nonumber
\end{align}
In the second line of the above equation, we used $f_{\rm{min}} \propto T_{\rm{obs}}^{-3/8}$. In total, $\Gamma_{ab} \propto T_{\rm{obs}}^{31/8}$, which leads to $\Delta w_0,\, \Delta w_a \propto T_{\rm{obs}}^{-31/16}$. Note that the $r_n$ dependence in Eq.~(\ref{eq21}) is not a good approximation for large $T_{\rm{obs}}$ because of the presence of the WD foreground. For the same reason, the deviation of the above scaling for $T_{\rm{obs}}$ is also large for small $r_n$. Thus, the scaling relations in $\Delta w_0$ and $\Delta w_a$ should be considered as qualitative one.

To show the results more clearly, let us define the figure of merit (FoM) for dark energy,
\begin{equation}
{\rm{FoM}} \equiv \sqrt{\det \gamma_{ab}} \;,
\label{eq30}
\end{equation}  
where $\gamma_{ab}$ is the inverse matrix of $(\Gamma^{-1})_{ab}$ with $a,b=w_0,\,w_a$. As is obvious from the definition, FoM is inversely proportional to the area of an error ellipse on the $w_0$ - $w_a$ plane. The FoM is listed in Table \ref{tab3} and is plotted in Fig.~\ref{fig5} as a function of $r_n$ for various values of $\dot{n}_0$ and $T_{\rm{obs}}$. From the scaling of $\Delta w_0$ and $\Delta w_a$, FoM is expected to scale as ${\rm{FoM}} \propto \dot{n}_0\, T_{\rm{obs}}^{31/8} r_n^{-2}$. 

\subsection{Redshift dependence of measurement accuracy}
\begin{figure}[t]
\begin{center}
\includegraphics[width=8cm]{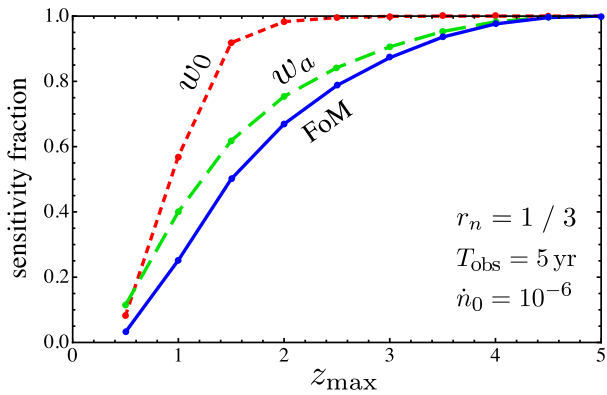}
\caption{Redshift dependence of the measurement accuracies in $d_L$ - $X$ measurement: ${\rm{FoM}}/{\rm{FoM}}(z_{\rm{max}}=5)$ (blue, solid), $\Delta w_0(z_{\rm{max}}=5)/\Delta w_0$ (red, dotted), $\Delta w_a(z_{\rm{max}}=5)/\Delta w_a$ (green, dashed). Parameters are fixed to $r_n=1/3$, $T_{\rm{obs}}=5\,{\rm{yr}}$, and $\dot{n}_0=10^{-6}\,{\rm{Mpc}}^{-3}\,{\rm{yr}}^{-1}$.}
\label{fig12}
\end{center}
\end{figure}

The measurement accuracies in the previous subsection are calculated with NS binaries detected up to $z=5$. From an observational point of view, it is interesting to see which redshift sources largely contribute to the measurement accuracies. To do this, we fix the parameters to $r_n=1/3$, $T_{\rm{obs}}=5\,{\rm{yr}}$, and $\dot{n}_0=10^{-6}\,{\rm{Mpc}}^{-3}\,{\rm{yr}}^{-1}$ and compute the measurement errors and FoMs, restricting the redshift range of the binary distribution below $z_{\rm{max}}=0.5 \times i$ ($i=1,2,\cdots,10$), i.e. upper cutoff of the integral in Eq.~(\ref{eq12}). The results are shown in Fig.~\ref{fig12} in terms of the sensitivity fractions: ${\rm{FoM}}/{\rm{FoM}}(z_{\rm{max}}=5)$, $\Delta w_0(z_{\rm{max}}=5)/\Delta w_0$, and $\Delta w_a(z_{\rm{max}}=5)/\Delta w_a$ as a function of $z_{\rm{max}}$, where say, $\Delta w_0(z_{\rm{max}}=5)$ represents the $w_0$ error with full redshift range. 

The figure shows that the sensitivity fractions sharply increase up to $z=1.5 - 2.0$ and slowly approach the maximum value at $z=5$. This reflects merely the redshift distribution of NS binaries in Eq.~(\ref{eq1}). In other words, the brighter source at low $z$ gives large SNR but the number of the source is small. A large number of sources at modest $z$ largely contributes the measurement accuracies of cosmological parameters. The interesting feature is that the slope of $\Delta w_a(z_{\rm{max}}=5)/\Delta w_a$ is gradual, compared with $\Delta w_0(z_{\rm{max}}=5)/\Delta w_0$, because the samples at high $z$ are required to constrain the time variation of dark energy, $w_a$ (see Fig.~\ref{fig4}). The fraction ${\rm{FoM}}/{\rm{FoM}}(z_{\rm{max}}=5)$ traces the redshift dependence of the $w_a$ error and shows gradual increase.

\subsection{Adding redshift information of binaries}
In the previous subsections, we used the combination of the luminosity distance $d_L$ and the cosmological phase correction $X$ to measure the dark-energy parameters. With this method, the cosmological expansion can be probed by GW observation alone without the identification of binary redshifts. Although it would be too ideal to assume that all binary redshifts are well determined by electromagnetic follow-up observations, some fraction of them will be determined by the follow-up observations. Then, we can optimize the constraints on the dark-energy parameters by using thus-determined redshifts. In this subsection, we combine $z$ - $d_L$ and $d_L$ - $X$ information and evaluate what fraction is needed to achieve ${\rm{FoM}}=100$, which is a typical value expected in future dark energy surveys \cite{bib71}. 

The relevant Fisher matrix is defined with the redshift-identification fraction $\alpha$ by
\begin{align}
\Gamma_{ab} &= \alpha \int_0^{\infty} \frac{\partial_a d_L (z) \partial_b d_L(z)}{\sigma_{d_L}^2(z)} \frac{d N(z)}{dz} dz \nonumber \\
&+ \int_0^{\infty} \frac{\partial_a X(z) \partial_b X(z)}{\sigma_X^2(z)} \frac{d N(z)}{dz} dz \;, 
\label{eq14}
\end{align}
where $\sigma_{d_L}$ is an error in the luminosity distance, given in Eq.~(\ref{eq22}) in Appendix \ref{app2}. The free parameters of the Fisher marix are again $w_0$, $w_a$, and $\Omega_{\rm{m}}$, while the Hubble constant is fixed. Strictly speaking, the variables $d_L$ and $X$ cannot be treated independently as in Eq.~(\ref{eq14}). However, the correlation coefficient between $d_L$ and $X$ is of the order of $10^{-6}$. This reflects the fact that amplitude and phase are independent quantities, hence we can safely use the expression in Eq.~(\ref{eq14}). The fraction factor $\alpha$ would be a function of a redshift in a real galaxy survey, but we take it as a constant for simplicity of the analysis.

In Appendix \ref{app2}, we estimate the error size of the luminosity distance via the same procedure as in Sec.~\ref{sec5b}, but including possible systematic errors: weak-lensing magnification due to the matter inhomogeneities along the line of sight and the peculiar velocity of each binary source. Then substituting the estimated errors for Eq.~(\ref{eq14}), we calculate $\alpha$ needed to achieve ${\rm{FoM}}=100$, which we denote by $\alpha_{100}$, and the corresponding number of NS binaries $N_z$ whose redshifts are to be identified. 

\begin{figure}[t]
\begin{center}
\includegraphics[width=8cm]{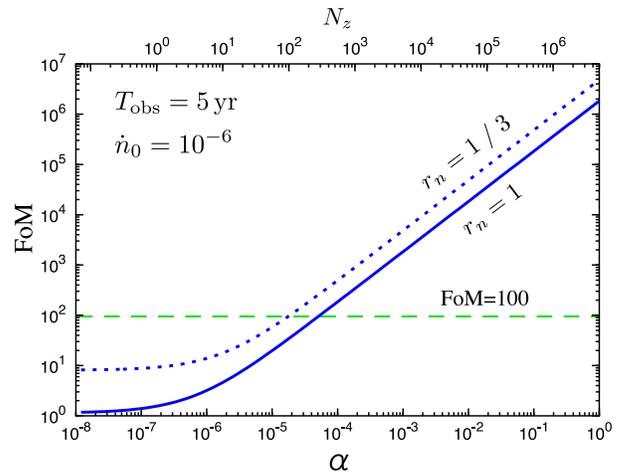}
\caption{FoM as a function of $\alpha$ defined in Eq.~(\ref{eq14}) or $N_z$. The observation time and the merger rate are fixed to $T_{\rm{obs}}=5\,{\rm{yr}}$ and $\dot{n}_0=10^{-6}\,{\rm{Mpc}}^{-3}\,{\rm{yr}}^{-1}$. The horizontal dashed line represents ${\rm{FoM}}=100$. The solid and dotted curves are the case with $r_n=1$ and $1/3$, respectively.}
\label{fig11}
\end{center}
\end{figure}

Before showing the results, it would be helpful to see how FoM changes as $\alpha$ increases. In Fig.~\ref{fig11}, FoM as a function of $\alpha$ is presented for fixed $T_{\rm{obs}}$ and $\dot{n}_0$. At low $\alpha$, of course, the $z$ - $d_L$ information plays no significant role in improving the FoM. But at $\alpha \approx$ a few $\times10^{-6}$, $z$ - $d_L$ information begins to contribute to FoM, and enables the FoM to reach 100 at around $\alpha \approx$ a few $\times 10^{-5}$. The fraction $\alpha_{100}$ required to achieve ${\rm{FoM}}=100$ is listed in Table~\ref{tab5}. As anticipated, $z$ - $d_L$ measurement is much more powerful than the $d_L$ - $X$ measurement because ${\rm{FoM}}|_{\alpha=0}$ is much smaller than ${\rm{FoM}}|_{\alpha=1}$. However, how much fraction $\alpha$ is feasible in the future galaxy redshift surveys? In the next subsection, we estimate $\alpha$ from rough consideration based on the number density of galaxies potentially observable and the number of galaxies to be observed in the future galaxy redshift surveys.

\begin{table*}[t]
\begin{center}
\begin{tabular}{|c|c|c|c|c|c|c|}
\hline 
\multicolumn{7}{|c|}{$T_{\rm{obs}}=3\,{\rm{yr}}$} \\
\hline 
& \multicolumn{2}{|c|}{$\dot{n}_0=10^{-7}$} & \multicolumn{2}{|c|}{$\dot{n}_0=10^{-6}$} & \multicolumn{2}{|c|}{$\dot{n}_0=10^{-5}$} \\
\hline 
$\quad r_n \quad$ & $\quad \quad \alpha_{100} \quad \quad$ & $\quad N_z \quad$ & $\quad \quad \alpha_{100} \quad \quad$ & $\quad N_z \quad$ & $\quad \quad \alpha_{100} \quad \quad$ & $\quad N_z \quad$ \\
\hline
$1$ & $9.1\times 10^{-4}$ & 270 & $9.3\times 10^{-5}$ & 280 & $1.1\times10^{-5}$ & 330 \\ 
$1/2$ & $4.4\times 10^{-4}$ & 130 & $4.6\times 10^{-5}$ & 140 & $4.0\times10^{-6}$ & 120 \\ 
$1/3$ & $3.5\times 10^{-4}$ & 100 & $3.4\times 10^{-5}$ & 100 & $2.8\times10^{-6}$ & 80 \\ 
$1/5$ & $3.0\times 10^{-4}$ & 90 & $2.9\times 10^{-5}$ & 90 & $1.7\times10^{-6}$ & 50 \\ 
\hline
\multicolumn{7}{c}{} \\
\hline
\multicolumn{7}{|c|}{$T_{\rm{obs}}=5\,{\rm{yr}}$} \\
\hline 
& \multicolumn{2}{|c|}{$\dot{n}_0=10^{-7}$} & \multicolumn{2}{|c|}{$\dot{n}_0=10^{-6}$} & \multicolumn{2}{|c|}{$\dot{n}_0=10^{-5}$} \\
\hline 
$r_n$ & $\alpha_{100}$ & $N_z$ & $\alpha_{100}$ & $N_z$ & $\alpha_{100}$ & $N_z$ \\
\hline
$1$ & $5.3\times 10^{-4}$ & 260 & $5.4\times 10^{-5}$ & 270 & $5.7\times10^{-6}$ & 280 \\ 
$1/2$ & $2.6\times 10^{-4}$ & 130 & $2.5\times 10^{-5}$ & 120 & $1.4\times 10^{-6}$ & 70 \\ 
$1/3$ & $2.1\times 10^{-4}$ & 100 & $1.9\times 10^{-5}$ & 100 & $3.0\times 10^{-7}$ & 20 \\ 
$1/5$ & $1.8\times 10^{-4}$ & 90 & $1.4\times 10^{-5}$ & 70 & $0$ & 0 \\ 
\hline
\multicolumn{7}{c}{} \\
\hline
\multicolumn{7}{|c|}{$T_{\rm{obs}}=10\,{\rm{yr}}$} \\
\hline 
& \multicolumn{2}{|c|}{$\dot{n}_0=10^{-7}$} & \multicolumn{2}{|c|}{$\dot{n}_0=10^{-6}$} & \multicolumn{2}{|c|}{$\dot{n}_0=10^{-5}$} \\
\hline 
$r_n$ & $\alpha_{100}$ & $N_z$ & $\alpha_{100}$ & $N_z$ & $\alpha_{100}$ & $N_z$ \\
\hline
$1$ & $2.6\times 10^{-4}$ & 260 & $2.2\times 10^{-5}$ & 220 & $0$ & 0 \\ 
$1/2$ & $1.2\times 10^{-4}$ & 120 & $6.4\times 10^{-6}$ & 60 & $0$ & 0 \\ 
$1/3$ & $9.4\times 10^{-5}$ & 90 & $1.5\times 10^{-6}$ & 20 & $0$ & 0 \\ 
$1/5$ & $7.4\times 10^{-5}$ & 70 & $0$ & 0 & $0$ & 0 \\ 
\hline
\end{tabular}
\end{center}
\caption{The fraction $\alpha$ required to reach ${\rm{FoM}}=100$ and the corresponding number of a NS binary that needs redshift identification. The binary number is rounded off to tens. $\alpha_{100}=0$ or $N_z=0$ in the table means that ${\rm{FoM}}=100$ is already achieved by $X$ measurement without any redshift information. $\dot{n}_0$ is in the unit of ${\rm{Mpc}}^{-3}\,{\rm{yr}}^{-1}$.}
\label{tab5}
\end{table*}

\subsection{On the redshift determination from optical follow-up observations}
\label{sec5E}
To obtain the redshift of a binary source, we need to identify its host galaxy and determine the redshift by an electromagnetic follow-up observation. As shown by Cutler and Holz \cite{bib4} from the simple consideration based on the average number density of galaxies, DECIGO and BBO has the angular resolution $\sim 1-100\,{\rm{arcsec}}^2$ and can uniquely identify the host galaxy of the binary. Although more complication such as galaxy clustering and so on should be taken into account, in what follows, we assume that a single host galaxy is found in a detector error cube. So our main concern here is what fraction of GW sources can we determine the redshifts by electromagnetic follow-up observations in a realistic situation with finite observation time and limited survey magnitude. 

The number density of galaxies potentially observable (brighter than a limiting apparent magnitude) can be estimated from the galaxy luminosity function. According to the paper of the Hubble Ultra Deep Field \cite{bib85}, the total number of galaxies potentially observable per unit square angle with the limiting magnitude $m_{\rm{lim}}\approx 29$ in an optical band is $\sim 5\times 10^3\,/{\rm{arcmin}}^2$. The total number over the celestial sphere is $\sim 7\times 10^{11}$. However, in a real observation, various factors such as assigned observation time and accessible observation bands restrict the number of galaxies to be observed in the survey. The future galaxy spectroscopic surveys, JDEM/WFIRST \cite{bib86} and Euclid \cite{bib87}, plan to observe $\sim 10^8$ galaxies in the redshift range $0.5 < z < 2$ with the redshift precision better than $0.1\%$ \cite{bib84}. Hence, the fraction of the galaxies whose redshifts are listed in a galaxy catalog is $f_{\rm{catalog}} \sim 1\times 10^{-4}$. For simplicity, we assume that GW events randomly occur in one of the galaxies. Unless we arrange follow-up observations especially dedicated for the GW events, the probability that the host galaxy of a GW event is listed in the galaxy catalog is also $f_{\rm{catalog}}$, which gives the fraction of redshift determination of binary sources.

The estimation based on the galaxy catalog is simple but seems to be reasonable, because galaxy redshift surveys are efficient in that they first select the candidate galaxies for the spectroscopic observation by luminosity and emission lines. This means that other galaxies not listed in the catalog is more difficult to obtain the redshift. So we expect that the number of redshift-determined binary sources would not be significantly changed even if we do arrange follow-up observations targeted at the GW events.

On the other hand, in the above estimate, it may be too simple to assume that the probability is independent of the mass and type of the galaxy and that GW events randomly occur in one of the galaxies. In fact, GW events are likely to occur in more massive luminous galaxies, and the fraction of redshift determination of binary sources would be larger. In that sense, our estimate can be considered as the worst case and contains large uncertainty.  

Another way to determine the source redshift is to identify the electromagnetic counterpart when a NS binary merges, which is believed to be the short gamma-ray burst (GRB). From GW observations in DECIGO band, we would be able to alert the electromagnetic telescopes to the event typically a few years before the merger. Given the half opening angle of a short GRB is $\approx 10\,{\rm{deg}}$, about $2\%$ of the short GRBs point toward us. Even if we assume that only $10\%$ of them have measurable redshift due to a noisy spectrum, dimming at high redshifts, and various telescope conditions, the fraction of redshift identification is $\approx 2\times 10^{-3}$, which is higher than that of the host galaxy identification. However, the opening angle of short GRBs has been poorly constrained \cite{bib96}. Clarifying whether this method provides adequate counterparts needs further observation and more statistical samples.

From the above estimations, it turns out that successful redshift identification seems
difficult and rather nontrivial. This indicates that the successful identification with high efficiency generally requires a dedicated follow-up mission with a sophisticated
strategy. Therefore, the standard sirens without source redshifts guarantee minimally achievable FoM and strongly supports the feasibility of the space-based GW detectors as a high-precision dark energy probe. 

Finally, note that the future galaxy surveys provide the redshifts of host galaxies below $z \approx 2$, while GW events distribute out to $z\approx 5$. However, from Eq.~(\ref{eq1}), most of GW events are localized around $z=1-2$. In addition, the high-$z$ GW sources are detected with less SNRs, and the existence of the dark energy would be less important at high-$z$. From these reasons, the follow-up observations at high-$z$, which is difficult and time-consuming, are not significantly crucial for the sensitivity of the standard siren.   

\section{Primordial gravitational waves}
\label{sec6}

Next we will move on to the sensitivity of DECIGO and BBO to a primordial GW background. To search for the primordial GW background, one needs to accurately identify the waveform of NS binaries and subtract them up to the level sufficiently below the amplitude of the primordial GW background. In that sense, the parameter determination of the NS binaries and the detection of primordial GW background are not independent topics but are closely related with each other.

\subsection{SNR formula}

Here we consider a stochastic GW background that is (i) isotropic, (ii) stationary, (iii) Gaussian, and (iv) unpolarized (see \cite{bib44} for the detailed discussions). To distinguish the GW-background signal from stochastic detector noise, one has to correlate signals between two detectors whose instrumental noises are uncorrelated with each other. The correlation-analysis technique has been well developed by several authors \cite{bib42,bib43,bib44} and has been extensively studied in the detector configuration of DECIGO and BBO \cite{bib45,bib46,bib47,bib48}. Under the assumptions that the energy density of the primordial GW background, $\Omega_{\rm{gw}}(f)$, is independent of frequency and that the amplitude of its GW signal is smaller than that of instrumental noise, the SNR formula for two interferometers located at the same position, which is the case of DECIGO and BBO, is written as
\begin{align}
{\rm{SNR}} &= \frac{9H_0^2}{20 \sqrt{2}\, \pi^2 } \sqrt{N_{\rm{corr}} T_{\rm{obs}}} \nonumber \\
&\times \biggl[ \int_{f_{\rm{min}}}^{f_{\rm{max}}} df \frac{\,\Omega ^2_{\rm{gw}} \Theta \left[ \Omega_{\rm{gw}}-\Omega_{\rm{gw}}^{\rm{NS}}(f)\, {\cal{R}}_{\rm{NS}} \right]} {f^6 \{S_{h}(f)\}^2} \biggr]
^{1/2}\;, 
\label{eq15} 
\end{align}
where the overlap reduction function is set to unity \cite{bib82}. The frequency range is set to $f_{\rm{min}}=0.2\,{\rm{Hz}}$ to escape from the WD confusion noise, and $f_{\rm{max}}=100\,{\rm{Hz}}$ as in the previous section. The power spectrum $S_h$ is given in  Eq.~(\ref{eq6}), but here it is multiplied by a factor $(1/\sqrt{5})^2 (\sqrt{3}/2)^2$ for the consistency of the definitions of the signal and noise in Eq.~(\ref{eq15}). The step function guarantees $\Omega_{\rm{gw}} > \Omega_{\rm{gw}}^{\rm{NS}}(f)\, {\cal{R}}_{\rm{NS}}$, which is a necessary condition for the clean subtraction of NS binaries. $N_{\rm{corr}}$ is the number of independent correlation signals whose overlap reduction function is unity. Since DECIGO and BBO have two clusters at the same location and each cluster is composed of two independent ($45^{\circ}$-rotated) interferometers \cite{bib49}, $N_{\rm{corr}}=2$.

\subsection{Results}
\begin{figure}[t]
\begin{center}
\includegraphics[width=8cm]{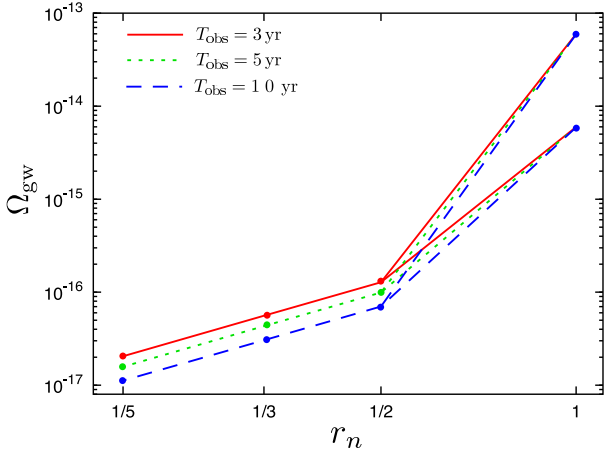}
\caption{$\Omega_{\rm{gw}}$ detectable with ${\rm{SNR}}=5$ for each parameter set of $r_n$, $T_{\rm{obs}}$, and $\dot{n}_0$. The observation time is denoted by colors: $T_{\rm{obs}}=3\,{\rm{yr}}$ (red, solid), $5\,{\rm{yr}}$ (green, dotted), $10\,{\rm{yr}}$ (blue, dashed). For a fixed observation time, the merger rate only affects the $r_n=1$ case: the upper and lower lines correspond to $\dot{n}_0=10^{-6}$, and $10^{-7}$ ${\rm{Mpc}}^{-3}\,{\rm{yr}}^{-1}$, respectively. For $\dot{n}_0=10^{-5}$ case, no data is plotted due to the same reason in Table \ref{tab6}.}
\label{fig10}
\end{center}
\end{figure}

Using Eq.~(\ref{eq15}), we calculate $\Omega_{\rm{gw}}$ detectable with ${\rm{SNR}}=5$. Table \ref{tab6} and Fig.~\ref{fig10} summarize the resultant $\Omega_{\rm{gw}}$ for various $T_{\rm{obs}}$ and $\dot{n}_0$. The sensitivity to $\Omega_{\rm{gw}}$ in all cases with $r_n=1/2,\,1/3,\,1/5$ are independent of the binary-merger rate $\dot{n}_0$, because all NS binaries are subtracted out. As for the $r_n=1$ case, some fraction of NS binaries, which is proportional to $\dot{n}_0$, limits the sensitivity to $\Omega_{\rm{gw}}$. For $\dot{n}_0=10^{-5}\,{\rm{Mpc}}^{-3}\,{\rm{yr}}^{-1}$, since a part of the noise curve is dominated by the NS-binary residual and the assumption that the amplitude of the GW background is smaller than the noise is no longer valid, the sensitivity based on Eq.~(\ref{eq15}) is not shown in Table \ref{tab6}. As explicitly understood from Eq.~(\ref{eq15}), the sensitivity to $\Omega_{\rm{gw}}$ is improved proportional to $r_n^2$ and $T_{\rm{obs}}^{-1/2}$ except for the case with $r_n=1$, where the sensitivity is limited by the NS-binary residual. 

\begin{table}[t]
\begin{center}
\begin{tabular}{|c|c|c|c|}
\hline
\multicolumn{4}{|c|}{$T_{\rm{obs}}=3\,{\rm{yr}}$} \\
\hline
$\quad r_n \quad$ & $\quad \dot{n}_0=10^{-7} \quad$ & $\quad \dot{n}_0=10^{-6} \quad$ & $\quad \dot{n}_0=10^{-5} \quad$ \\
\hline
$1$ & $5.97\times 10^{-15}$ & $5.97\times 10^{-14}$ & ----- \\ 
$1/2$ & $9.05\times 10^{-17}$ & $9.05\times 10^{-17}$ & $9.05\times 10^{-17}$ \\ 
$1/3$ & $4.02\times 10^{-17}$ & $4.02\times 10^{-17}$ & $4.02\times 10^{-17}$ \\ 
$1/5$ & $1.45\times 10^{-17}$ & $1.45\times 10^{-17}$ & $1.45\times 10^{-17}$ \\ 
\hline
\multicolumn{4}{c}{} \\
\hline
\multicolumn{4}{|c|}{$T_{\rm{obs}}=5\,{\rm{yr}}$} \\
\hline
$r_n$ & $\dot{n}_0=10^{-7}$ & $\dot{n}_0=10^{-6}$ & $\dot{n}_0=10^{-5}$ \\
\hline
$1$ & $5.94\times 10^{-15}$ & $5.94\times 10^{-14}$ & ----- \\ 
$1/2$ & $7.01\times 10^{-17}$ & $7.01\times 10^{-17}$ & $7.01\times 10^{-17}$ \\ 
$1/3$ & $3.12\times 10^{-17}$ & $3.12\times 10^{-17}$ & $3.12\times 10^{-17}$ \\ 
$1/5$ & $1.12\times 10^{-17}$ & $1.12\times 10^{-17}$ & $1.12\times 10^{-17}$ \\ 
\hline
\multicolumn{4}{c}{} \\
\hline
\multicolumn{4}{|c|}{$T_{\rm{obs}}=10\,{\rm{yr}}$} \\
\hline
$r_n$ & $\dot{n}_0=10^{-7}$ & $\dot{n}_0=10^{-6}$ & $\dot{n}_0=10^{-5}$ \\
\hline
$1$ & $5.92\times 10^{-15}$ & $5.92\times 10^{-14}$ & ----- \\ 
$1/2$ & $4.96\times 10^{-17}$ & $4.96\times 10^{-17}$ & $4.96\times 10^{-17}$ \\ 
$1/3$ & $2.20\times 10^{-17}$ & $2.20\times 10^{-17}$ & $2.20\times 10^{-17}$ \\ 
$1/5$ & $7.93\times 10^{-18}$ & $7.93\times 10^{-18}$ & $7.93\times 10^{-18}$ \\ 
\hline
\end{tabular}
\end{center}
\caption{Sensitivity to $\Omega_{\rm{gw}}$ (${\rm{SNR}}=5$). $\dot{n}_0$ is in the unit of ${\rm{Mpc}}^{-3}\,{\rm{yr}}^{-1}$. The symbol "---" in the table indicates that the NS residual partially covers the instrumental-noise curve and the SNR formula in Eq.~(\ref{eq15}) cannot be used to compute the sensitivity.}
\label{tab6}
\end{table}

\section{Discussions}
\label{sec7}

\subsection{Impact of GW observations on cosmology}

\begin{figure*}[t]
\begin{center}
\includegraphics[width=17cm]{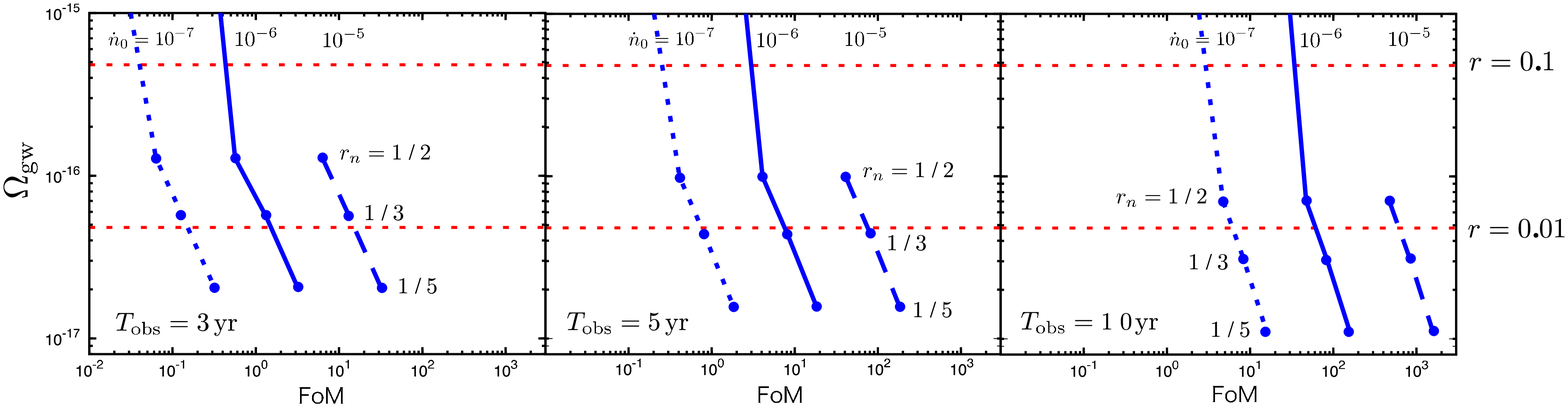}
\caption{FoM and $\Omega_{\rm{gw}}$ achievable with GW observation alone ($\eta=0$) for each parameter set of $r_n$, $T_{\rm{obs}}$, and $\dot{n}_0$. From the left to the right panels, the observation time is $3\,{\rm{yr}}$, $5\,{\rm{yr}}$, and $10\,{\rm{yr}}$. In each panel, the merger rate is represented by curve types: $\dot{n}_0=10^{-5}$ (dashed), $10^{-6}$ (solid), $10^{-7}$ (dotted), in the unit of ${\rm{Mpc}}^{-3}\,{\rm{yr}}^{-1}$. On each curve, points from the top to the bottom correspond to $r_n=1/2$, $1/3$, and $1/5$. The horizontal lines correspond to $r=0.1$ and $0.01$ at CMB scale in the de-Sitter inflation.}
\label{fig7}
\end{center}
\end{figure*}

In the previous sections, we presented the sensitivity to dark energy and a GW background of a space-based detectors, which are summarized in Fig.~\ref{fig7}. In what follows, we discuss some scientific consequences and the detector sensitivity. 

From $d_L$ - $X$ measurement, FoM for dark energy computed in Sec.~\ref{sec5} ranges from $\sim 0.1$ to $\sim 10^3$, depending on the observation time, the merger rate, and the instrumental-noise level. Since DECIGO and BBO will be launched in the late 2020s, the FoM should be compared with other future projects of an electromagnetic observation probing dark energy at that time: type-Ia supernovae, baryon acoustic oscillation, or weak-lensing surveys. So typical criteria in FoM would be from 10 to 100, which correspond to future projects of stage III and IV in the dark energy task force \cite{bib71}, respectively. To achieve these criteria, observation time longer than $5\,{\rm{yr}}$ is preferable. Since the FoM rapidly improves being roughly proportional to $T_{\rm{obs}}^{31/8}$, the observation time is a crucial factor. Given 10-yr observation and the typical rate of binary mergers, the FoM is $\sim 100$ and is comparable to a stage-IV project in the dark energy task force. It should be emphasized that this method requires no redshift information of the GW sources and is completely independent of any electromagnetic observation. Of course, if the GW observation data are combined with those of other electromagnetic observation, the FoM will be much improved as in the case of type-Ia supernovae. On the other hand, identifying the source redshifts strongly assists the $d_L$ - $X$ measurement. If the fraction of source-redshift determination $\alpha$ is larger than $10^{-4}$ in the case with $5\,{\rm{yr}}$-observation and $\dot{n}_0=10^{-6}\,{\rm{Mpc}}^{-3}\,{\rm{yr}}^{-1}$, the FoM is mainly determined by the binary sources with redshift information. The order-of-magnitude estimate of the fraction of redshift determination could be about $10^{-4}$ in the worst case. Hence, we can conclude that the standard sirens without source redshifts guarantee minimally achievable FoM. 

As for a GW background, as shown in Table \ref{tab6}, the space-based GW detectors such as DECIGO and BBO have enough sensitivity to directly probe a primordial GW background. On the other hand, cosmic microwave background (CMB)-polarization experiments \cite{bib51,bib52,bib53,bib54} are also sensitive to the primordial GW background at the cosmological-horizon scale at present. Although the frequency range is far apart from that of the direct-detection experiment, the detector sensitivities should be compared with one another if one targets the detection of inflationary GWs. This is because the spectrum of GW-background energy density $\Omega_{\rm{gw}}$ generated by single-field slow-roll inflation is predicted to be red-tilted \cite{bib55,bib56,bib57} and is related to the tensor-to-scalar ratio defined at CMB scale via the evolution of the slow-roll parameters as \cite{bib62,bib63} 
\begin{equation}
\Omega_{\rm{gw}} (f_*) = 4.81\times 10^{-15} h_{72}^{-2}\, r \left( \frac{f_*}{f_0} \right)^{n_T} \;. \nonumber
\end{equation}
Here $f_0=3.24\times10^{-18}\,{\rm{Hz}}$, and $f_*$ is the frequency of the GW detector, e.g. $0.1 - 1 \,{\rm{Hz}}$. $n_T$ is the spectral index of the tensor mode. 
Although the tilt and running of the spectrum have not been observationally well-constrained, the upper bound on the magnitude of the GW spectrum in the direct-detection frequency band is obtained in the limit of de-Sitter inflation ($n_T=0$) as
\begin{equation}
\Omega_{\rm{gw}}(f_*) = 4.81\times 10^{-15}\, r \;. \nonumber
\end{equation}
Here we set $h_{72}=1$. Therefore, if the inflationary GWs are not detected by the CMB experiment, the GW detectors at least should have better sensitivity than the preceded CMB experiments for the first detection of an slow-roll inflationary GWs \cite{bib83}. Current constraint on the tensor-to-scalar is $r < 0.2$ \cite{bib50}, which is translated into $\Omega_{\rm{gw}} < 9.6 \times 10^{-16}$ at $f=f_*$. The future CMB-polarization experiments would be able to achieve $r=0.01$, corresponding to $\Omega_{\rm{gw}} = 4.8 \times 10^{-17}$ at $f=f_*$. To surpass the CMB-experiment sensitivities, $r_n \leq 1/3$ is required for the noise spectrum of the GW detector. Note that the current preconceptual design of the BBO noise curve marginally satisfies this sensitivity requirement.
 
\subsection{Other potential systematics}
\label{sec7c}
In this paper so far, we consider systematic errors in the luminosity distance $d_L$ but not in the cosmological phase shift $X$. Since the cosmological phase shift is identical to a cosmological redshift drift, the signal could be contaminated due to the peculiar acceleration of a source relative to an observer and the change of gravitational potential along the line of sight. According to Ref. \cite{bib72}, such systematic noises are estimated, based on a linear perturbation theory. They found that the dominant contribution is the peculiar acceleration of the source, but the magnitude is smaller by one- or two- orders than the cosmological signal in the typical range of a redshift, i.e. $z \approx 0.1 - 3$. In Refs. \cite{bib73,bib74}, taking the nonlinear effect of density perturbations into account, the authors considered a source in a galaxy and a galaxy cluster. In this case, the peculiar acceleration of the source is of the same order as the cosmological signal for a single source. However, the systematic error would be random for each source and can be reduced by averaging it out. Therefore, for our observation, we can conclude that the peculiar acceleration does not contribute to the cosmological phase shift of a GW.

Another possible error that affects the sensitivity to a primordial GW background is subtraction noise that comes from parameter mismatches of each NS binary in the subtraction process. Although we neglect this contribution in this paper, it should be considered seriously because clean subtraction is crucial for detecting a primordial GW background. Such a numerical investigation in the BBO detector configuration has been performed by Harms {\it{et al.}} \cite{bib90}. They showed that the residual noise is significantly suppressed by the subtraction process with the projection method and that BBO is able to detect the GW background with $\Omega_{\rm{gw}} \approx 2.5 \times 10^{-17}$. Their result is obtained in the situation that $\sim 10^5$ NS binaries are detected during 3-yr observation. Longer observation time would lead to the same conclusion, because the projected spectrum decrease with $1/T_{\rm{obs}}$. However, they assumed the pessimistic merger rate, $\dot{n}_0=10^{-7}\,{\rm{Mpc}}^{-3}\,{\rm{yr}}^{-1}$, which is 10 times smaller than the rate we used in this paper. If the rate is larger, the applicability of their result should be reconsidered. Since the BBO sensitivity is comparable to DECIGO with $r_n=1/3$, it is obscure how DECIGO with worse sensitivity affects the result. This issue of parameter mismatches should be investigated further in more broad parameter range in the future work. 

\subsection{Signal overlaps}
DECIGO will detect $\sim 10^6$ of NS binaries, whose signals could overlap and become individually indistinguishable. To apply the projection method for the parameter-mismatch noise mentioned in the previous subsection, we at least need to subtract nearly matched signals. In this subsection, we discuss under what conditions these overlaps cause a problem in separately extracting each signal. 

The number of inspiral GW signals $\Delta N (f)$ in a bin of minimum frequency resolution $\Delta f = 1/T_{\rm{obs}}$ is given by
\begin{equation}
\Delta N (f) = \frac{dN}{dt} \left( \frac{df}{dt} \right)^{-1} \Delta f \;. 
\label{eq31} 
\end{equation}  
Here $dN/dt$ is the merger rate per unit time, which is roughly 
\begin{equation}
\frac{dN}{dt} \approx 10^6\,{\rm{yr}}^{-1} \left( \frac{\dot{n}_0}{10^{-6}\,{\rm{Mpc}}^{-3}\,{\rm{yr}}^{-1}} \right) \;. \nonumber 
\end{equation} 
Substituting the frequency derivative \cite{bib18,bib19}
\begin{equation}
\frac{df}{dt} = \frac{96}{5} \pi^{8/3} M_z^{5/3} f^{11/3} \;, \nonumber
\end{equation}
into Eq.~(\ref{eq31}) and setting $\Delta N (f_{\rm{conf}})=1$, we obtain the critical frequency below which more than two signals are in the same frequency bin as
\begin{equation}
f_{\rm{conf}} \approx \left( \frac{5}{96\, T_{\rm{obs}}} \frac{dN}{dt} \right)^{3/11} \pi^{-8/11} M_z^{-5/11} \;. \nonumber
\end{equation}
Roughly speaking, below this frequency, each signal is not distinguishable, which prevents us from extracting it individually. 

If we assume $T_{\rm{obs}}=5\,{\rm{yr}}$ and that all signals are at $z=1$ for simplicity, the frequency $f_{\rm{conf}}$ is $0.04\,{\rm{Hz}}$ for $\dot{n}_0 =10^{-7}\,{\rm{Mpc}}^{-3}\,{\rm{yr}}^{-1}$, $0.07\,{\rm{Hz}}$ for $\dot{n}_0 =10^{-6}\,{\rm{Mpc}}^{-3}\,{\rm{yr}}^{-1}$, and $0.14\,{\rm{Hz}}$ for $\dot{n}_0 =10^{-5}\,{\rm{Mpc}}^{-3}\,{\rm{yr}}^{-1}$. As seen from Fig.~\ref{fig2}, the sensitivity to dark energy is hardly affected for the first two choices of $\dot{n}_0$. However, if $\dot{n}_0 =10^{-5}\,{\rm{Mpc}}^{-3}\,{\rm{yr}}^{-1}$ and the lower cutoff frequency is set to $f_{\rm{min}}=0.2\,{\rm{Hz}}$, the FoMs are considerably deteriorated from $9.2$ to $6.6\times 10^{-3}$ for $r_n=1$ and from $81$ to $7.5\times 10^{-2}$ for $r_n=1/3$. These results are expected from the scaling relation ${\rm{FoM}} \propto T_{\rm{obs}}^{31/8}$, since $f_{\rm{min}}=0.2\,{\rm{Hz}}$ corresponds to about $0.35\,{\rm{yr}}$ before the merger. To avoid this overlap problem, one needs more ingenious analysis method. We will leave it as a future work. As for the detection of an inflationary GW background, since we have already set $f_{\rm{min}}=0.2\,{\rm{Hz}}$ in our analysis, the signal overlap does not directly affect the SNR. The degradation of the SNRs for NS binaries might increase the residual foreground though the parameter mismatches discussed in Sec.~\ref{sec7c}. However, if they are projected out \cite{bib90}, the residual would not affect the sensitivity.

\subsection{Comparison with other methods of the standard siren without redshifts}
There are two other proposals of the standard siren without electromagnetic counterparts, in which the degeneracy between a source redshift and a chirp mass is broken by observing a tidal effect on a GW \cite{bib89} and by constraining the redshift range of each binary source with the observed chirp mass distribution of NSs \cite{bib92}. In this subsection, we briefly comment on the comparison of our method with others. 

In the method utilizing the tidal effect \cite{bib89}, we need to {\it{a priori}} know the equation of state of a NS. Furthermore, the tidal correction on the GW phase is a 5 PN-order effect and can be observed only around the frequencies slightly before the merger, e.g. $\sim \,{\rm{kHz}}$. Thus, this method would be available for Einstein Telescope (ET) \cite{bib94} and is completely complementary to our method. It will be interesting to investigate the multiplier effect when NS binaries are observed in both low and high frequency ranges. In the method with a chirp mass distribution \cite{bib92}, the sensitivity strongly depends on the width of the distribution. The authors found that in order to measure $H_0$ with $\sim 10\,\%$ accuracy by advanced LIGO \cite{bib58}, assuming $\sim 100$ sources detected, the half-width of the mass distribution should be within $0.04\,M_{\odot}$. This method is also applicable to ET and space-based detectors. It is valuable to compare the sensitivities of these methods with that of our method, but it is beyond the scope of this paper and we leave it as a future work.

\section{Conclusion}
\label{sec8}

Proposed space-based GW detectors, DECIGO and BBO, can be a unique probe for dark energy and a primordial GW background. We investigated these target sciences from integrated point of view, allowing their noise curves to be scaled appropriately and including confusion noises produced by astrophysical sources. Using millions of NS binaries detected during the observation, we estimate the sensitivity to the parameters of the equation of state of the dark energy with/without identifying the redshifts of host galaxies. As a result, we found that the detectors without the redshift information of the sources have a constraining power competitive to the future electromagnetic observations. This is a great advantage since the GW detector alone can probe for the cosmological expansion and enables us to compare the data of purely GW observation with those obtained in other electromagnetic observations. With the help of the redshift information, ${\rm{FoM}}\approx 100$ corresponding to the stage III or IV of the dark energy task force is easily achieved with at most a few hundreds of sources. The standard sirens without redshifts guarantee minimally achievable FoM and strongly supports the feasibility of the space-based GW detectors as a high-precision dark energy probe. The detection and cleaning of GWs from the NS binaries are also necessary to search for a GW background generated in the early universe. We took the subtraction procedure of the NS binaries into account and computed the sensitivity to the amplitude of a GW background. As a result, the default BBO sensitivity or the DECIGO with three-times-better sensitivity is marginal in order to successfully subtract the NS foreground and to surpass the future CMB-experiment sensitivity to a primordial GW background. In summary, the subtraction of NS foreground is significantly important for both dark energy and primordial GW background, and our study here provides useful information to optimize the experiment.

\begin{acknowledgments}
We acknowledge S. Saito for 
helpful discussions and the anonumous referee for valuable comments to improve the manuscript. A. N. and K. Y. are supported 
by a Grant-in-Aid through JSPS. K. Y. is also supported in part by the Grant-in-Aid for the Global COE Program from the MEXT of Japan. A. T. is 
supported in part by a Grants-in-Aid for Scientific Research from 
the JSPS No. 21740168. 
\end{acknowledgments}

\appendix

\section{Fitting formula for NS residual}
\label{app1}
The NS residual fraction ${\cal{R}}_{\rm{NS}}$ as a function of $\rho_{\rm{th}}/\bar{\rho}_{\rm{e}}(5)$ is given by Yagi and Seto in \cite{bib17}. For the convenience of the analysis in this paper, we fit their result and provide the fitting formula \cite{bib81}. Fitting ${\cal{R}}_{\rm{NS}}$ by polynomials up to 5th order in $x \equiv \rho_{\rm{th}}/\bar{\rho}_{\rm{e}}(5)$ leads to the following piecewise expression:
\begin{widetext}
\begin{equation}
{\cal{R}}_{\rm{NS}} \approx \left\{ 
\begin{array}{lllll} 
-3.695+2.297\times10^1\,x -5.714\times10^1\,x^2 + 7.105\times10^1 \,x^3 -4.417\times10^1 \,x^4 + 1.098\times10^1 \,x^5 \\
\qquad \qquad \qquad \qquad \qquad \qquad \qquad \qquad \qquad \qquad \qquad \qquad \qquad \qquad \qquad \qquad \qquad {\rm{for}} \;\; x < 0.83 \\
9.675\times10^{-2}-5.663\times10^{-1} \,x +1.321 \,x^2 - 1.533 \,x^3 +8.847\times10^{-1} \,x^4 - 2.029\times10^{-1} \,x^5 \\
\qquad \qquad \qquad \qquad \qquad \qquad \qquad \qquad \qquad \qquad \qquad \qquad \qquad \qquad \qquad \qquad \qquad {\rm{for}} \;\;  0.83 \leq x < 0.9 \\
\\
2.166\times10^{-2}-1.255\times10^{-1} \,x +2.815\times10^{-1} \,x^2 - 3.039\times10^{-1} \,x^3 +1.563\times10^{-1} \,x^4 - 2.993\times10^{-2} \,x^5 \\
\qquad \qquad \qquad \qquad \qquad \qquad \qquad \qquad \qquad \qquad \qquad \qquad \qquad \qquad \qquad \qquad \qquad {\rm{for}} \;\; 0.9 \leq x < 1.1\\
\\
3.561\times10^{-4}-5.516\times10^{-3} \,x +2.321\times10^{-2} \,x^2 - 3.741\times10^{-2} \,x^3 +2.279\times10^{-2} \,x^4 - 4.424\times10^{-3} \,x^5 \\
\qquad \qquad \qquad \qquad \qquad \qquad \qquad \qquad \qquad \qquad \qquad \qquad \qquad \qquad \qquad \qquad \qquad {\rm{for}} \;\; 1.1 \leq x < 1.8 \\
\\
2.594\times10^{-2}-8.338\times10^{-2} \,x + 5.340\times10^{-2} \,x^2 -6.790\times10^{-3} \,x^3 +3.560\times10^{-4} \,x^4 + 6.722\times10^{-6} \,x^5 \\
\qquad \qquad \qquad \qquad \qquad \qquad \qquad \qquad \qquad \qquad \qquad \qquad \qquad \qquad \qquad \qquad \qquad {\rm{for}} \;\;  1.8 \leq x 
\end{array}
\right. \nonumber
\end{equation} 
\end{widetext}

\section{Comparing between error sizes of $d_L$ and $X$}
\label{app2}

In this Appendix, we estimate the error size of the luminosity distance $d_L$ in the presence of possible systematic errors and then compare it with that of cosmological phase correction $X$.

\begin{figure}[t]
\begin{center}
\includegraphics[width=8cm]{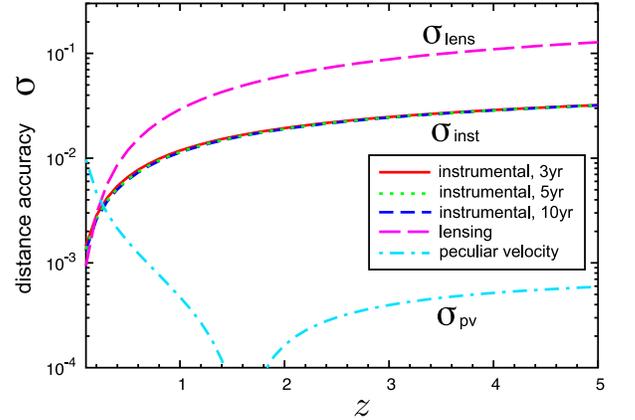}
\caption{Measurement accuracy of the luminosity distance with a single binary as a function of redshifts. The curves tagged $\sigma_{\rm{inst}}$ are those determined only by instrumental noise and with the observation time $3\,{\rm{yr}}$ (red, solid curve), $5\,{\rm{yr}}$ (green, dotted curve), and $10\,{\rm{yr}}$ (blue, short-dashed curve), respectively. The lensing error and the peculiar velocity error are represented by magenta (long-dashed) and light blue (dot-dashed) curves.}
\label{fig8}
\end{center}
\end{figure}

The error size of the luminosity distance can be estimated via the same procedure as in Sec.~\ref{sec5b}, but including systematic errors. It is known that the luminosity of a GW is magnified by the matter inhomogeneities of large-scale structure along the line of sight (e.g., \cite{bib38,bib39,bib33,bib40}), which systematically changes the luminosity distance to each binary system and contributes to the measured luminosity distance as a systematic error \cite{bib34}. In addition, the peculiar velocity of the binary along the line of sight randomly contributes to measurement error via Doppler effect \cite{bib36}.  
These systematic errors to the luminosity distance are summarized as 
\begin{align}
\sigma_{d_L}^2 (z) & \equiv \left[ \frac{\Delta d_L(z)}{d_L(z)} \right]^2 \nonumber \\
&= 
\sigma_{\rm{inst}}^2(z) + \sigma_{\rm{lens}}^2(z) + \sigma_{\rm{pv}}^2(z) \;,
\label{eq22} 
\end{align}
with
\begin{align}
\sigma_{\rm{lens}}(z) &=0.066 \left[ \frac{1-(1+z)^{-0.25}}{0.25} \right]^{1.8} 
\;, 
\label{eq19} \\
\sigma_{\rm{pv}}(z) &= \left| 1-\frac{(1+z)^2}{H(z)d_L(z)} \right| \sigma_{\rm{v,gal}} \;, \nonumber
\end{align}
where $\sigma_{\rm{inst}}$, $\sigma_{\rm{lens}}$, and $\sigma_{\rm{pv}}$ are induced by the instrumental noise, the lensing magnification, and the peculiar velocity of binaries, respectively. $\sigma_{\rm{v,gal}}$ is the one-dimensional velocity dispersion of the galaxy. Taking into account the nonlinear effect of gravity, it is often set to 
$\sigma_{\rm{v,gal}}=\,300$km\,s$^{-1}$, mostly independent of the redshifts 
\cite{bib41}. In Fig.~\ref{fig8}, the errors of the luminosity distance are illustrated. The lensing error dominates at almost all redshift range.

\begin{figure}[t]
\begin{center}
\includegraphics[width=7.5cm]{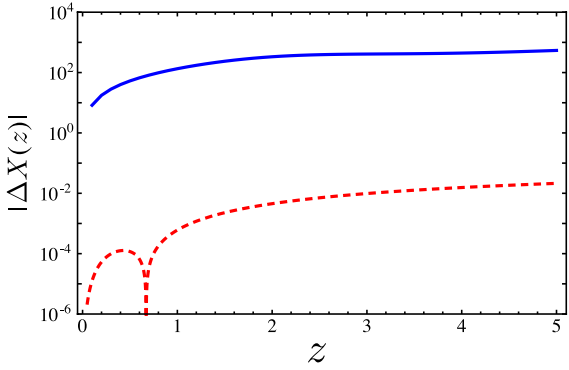}
\caption{Errors of $X(z)/H_0$ as a function of a redshift. The Solid curve denote the error due to instrumental noise in a GW observation, and the dotted curve denote the error propagating from the lensing error of $d_L$. Assumed parameters are $r_n=1$, $T_{\rm{obs}}=3\,{\rm{yr}}$, and $\dot{n}_0=10^{-6}\,{\rm{Mpc}}^{-3}\,{\rm{yr}}^{-1}$.} 
\label{fig9}
\end{center}
\end{figure}

We present $\Delta w_0$, $\Delta w_a$, and FoM  in the case with $\alpha=1$ in Table \ref{tab4}. As anticipated, $z$ - $d_L$ measurement is much more powerful than the $d_L$ - $X$ measurement.

\begin{table*}[t]
\begin{center}
\begin{tabular}{|c|c|c|c|c|c|c|c|c|c|}
\hline
\multicolumn{10}{|c|}{$T_{\rm{obs}}=3\,{\rm{yr}}$} \\
\hline
& \multicolumn{3}{|c|}{$\dot{n}_0=10^{-7}$} & \multicolumn{3}{|c|}{$\dot{n}_0=10^{-6}$} & \multicolumn{3}{|c|}{$\dot{n}_0=10^{-5}$} \\
\hline 
$r_n$ & $\Delta w_0$ & $\Delta w_a$ & FoM & $\Delta w_0$ & $\Delta w_a$ & FoM & $\Delta w_0$ & $\Delta w_a$ & FoM \\
\hline
$1$ & $7.833\times10^{-4}$ & $1.254\times 10^{-2}$ & $1.101\times10^{5}$ & $2.506\times10^{-4}$ & $4.007\times 10^{-3}$ & $1.078\times10^{6}$ & $8.696\times10^{-5}$ & $1.378\times10^{-3}$ & $9.095\times10^{6}$ \\ 
$1/2$ & $5.202\times10^{-4}$ & $8.782\times 10^{-3}$ & $2.272\times10^{5}$ & $1.692\times10^{-4}$ & $2.845\times 10^{-3}$ & $2.162\times10^{6}$ & $8.782\times10^{-4}$ & $4.632\times10^{-5}$ & $2.272\times10^{7}$ \\ 
$1/3$ & $4.537\times10^{-4}$ & $7.828\times 10^{-3}$ & $2.877\times10^{5}$ & $1.435\times10^{-4}$ & $2.475\times 10^{-3}$ & $2.877\times10^{6}$ & $4.537\times10^{-5}$ & $7.625\times10^{-4}$ & $2.877\times10^{7}$ \\ 
$1/5$ & $4.150\times10^{-4}$ & $7.265\times 10^{-3}$ & $3.357\times10^{5}$ & $1.312\times10^{-4}$ & $2.297\times 10^{-3}$ & $3.358\times10^{6}$ & $4.150\times10^{-5}$ & $7.265\times10^{-4}$ & $3.358\times10^{7}$ \\ 
\hline
\multicolumn{10}{c}{} \\
\hline
\multicolumn{10}{|c|}{$T_{\rm{obs}}=5\,{\rm{yr}}$} \\
\hline
& \multicolumn{3}{|c|}{$\dot{n}_0=10^{-7}$} & \multicolumn{3}{|c|}{$\dot{n}_0=10^{-6}$} & \multicolumn{3}{|c|}{$\dot{n}_0=10^{-5}$} \\
\hline 
$r_n$ & $\Delta w_0$ & $\Delta w_a$ & FoM & $\Delta w_0$ & $\Delta w_a$ & FoM & $\Delta w_0$ & $\Delta w_a$ & FoM \\
\hline
$1$ & $5.980\times10^{-4}$ & $9.611\times 10^{-3}$ & $1.875\times10^{5}$ & $1.913\times10^{-4}$ & $3.071\times 10^{-3}$ & $1.836\times10^{6}$ & $6.638\times10^{-5}$ & $1.056\times10^{-3}$ & $1.549\times10^{7}$ \\ 
$1/2$ & $3.997\times10^{-4}$ & $6.762\times 10^{-3}$ & $3.833\times10^{5}$ & $1.264\times10^{-4}$ & $2.138\times 10^{-3}$ & $3.833\times10^{6}$ & $3.997\times10^{-5}$ & $6.762\times 10^{-4}$ & $3.833\times10^{7}$ \\ 
$1/3$ & $3.498\times10^{-4}$ & $6.042\times 10^{-3}$ & $4.829\times10^{5}$ & $1.106\times10^{-4}$ & $1.911\times 10^{-3}$ & $4.829\times10^{6}$ & $3.498\times10^{-5}$ & $6.042\times 10^{-4}$ & $4.829\times10^{7}$ \\ 
$1/5$ & $3.213\times10^{-4}$ & $5.619\times 10^{-3}$ & $5.613\times10^{5}$ & $1.015\times10^{-4}$ & $1.777\times 10^{-3}$ & $5.613\times10^{6}$ & $3.208\times10^{-5}$ & $5.619\times 10^{-4}$ & $5.613\times10^{7}$ \\ 
\hline
\multicolumn{10}{c}{} \\
\hline
\multicolumn{10}{|c|}{$T_{\rm{obs}}=10\,{\rm{yr}}$} \\
\hline
& \multicolumn{3}{|c|}{$\dot{n}_0=10^{-7}$} & \multicolumn{3}{|c|}{$\dot{n}_0=10^{-6}$} & \multicolumn{3}{|c|}{$\dot{n}_0=10^{-5}$} \\
\hline 
$r_n$ & $\Delta w_0$ & $\Delta w_a$ & FoM & $\Delta w_0$ & $\Delta w_a$ & FoM & $\Delta w_0$ & $\Delta w_a$ & FoM \\
\hline
$1$ & $4.187\times10^{-4}$ & $6.746\times 10^{-3}$ & $3.803\times10^{5}$ & $1.339\times10^{-4}$ & $2.155\times 10^{-3}$ & $3.725\times10^{6}$ & $4.642\times10^{-5}$ & $7.405\times10^{-4}$ & $3.148\times10^{7}$ \\ 
$1/2$ & $2.813\times10^{-4}$ & $4.764\times 10^{-3}$ & $7.718\times10^{5}$ & $8.895\times10^{-5}$ & $1.507\times 10^{-3}$ & $7.718\times10^{6}$ & $2.813\times10^{-5}$ & $4.764\times 10^{-4}$ & $7.718\times10^{7}$ \\ 
$1/3$ & $2.468\times10^{-4}$ & $4.265\times 10^{-3}$ & $9.691\times10^{5}$ & $7.805\times10^{-5}$ & $1.349\times 10^{-3}$ & $9.691\times10^{6}$ & $2.468\times10^{-5}$ & $4.265\times 10^{-4}$ & $9.691\times10^{7}$ \\ 
$1/5$ & $2.267\times10^{-4}$ & $3.971\times 10^{-3}$ & $1.124\times10^{5}$ & $7.170\times10^{-5}$ & $1.256\times 10^{-3}$ & $1.124\times10^{6}$ & $2.267\times10^{-5}$ & $3.971\times 10^{-4}$ & $1.124\times10^{7}$ \\ 
\hline
\end{tabular}
\end{center}
\caption{Measurement accuracy of equation of state of dark energy and FoM when assumed that redshifts of all binaries are identified by electromagnetic follow-up observation of host galaxies ($\alpha=1$). $\dot{n}_0$ is in the unit of ${\rm{Mpc}}^{-3}\,{\rm{yr}}^{-1}$.}
\label{tab4}
\end{table*}

Next, we show that the error of $d_L$ is much smaller than that of $X$ and that $d_L$ can be regarded as a corresponding redshift in the fiducial model of the universe when we constrain cosmological parameters with standard sirens.

To compare the error sizes of $d_L$ and $X$, let us see whether the error of $d_L$ affects uncertainty of $X$ or not. For simplicity, we fix cosmological parameters to the fiducial values: $w_0=-1$, $w_a=0$, $\Omega_m=0.3$, $H_0=72\,{\rm{km}}\,{\rm{s}}^{-1}\,{\rm{Mpc}}^{-1}$ and regard $d_L$ and $X$ as a function of a redshift $z$. Differentiating Eqs.~(\ref{eq17}) and (\ref{eq11}) with respect to $z$ and eliminating $z$ give
\begin{equation}
\Delta X(z) = \frac{1}{2(1+z)} \frac{H-(1+z) \frac{dH}{dz}}{1+\frac{(1+z)^2}{d_L H}} \frac{\Delta d_L}{d_L} \;, 
\label{eq18}
\end{equation}
where
\begin{align}
\frac{d\,H(z)}{dz} &= \frac{1}{(1+z)H(z)} \nonumber \\
&\times \left[ \left( 1+w_0+w_a \frac{z}{1+z} \right) Q(z) +\frac{3}{2} H_0^2 \Omega_m (1+z)^3 \right] \;, \nonumber \\
& \nonumber \\
Q(z) &\equiv \frac{3}{2} H_0^2 (1-\Omega_m) (1+z)^{3(1+w_0+w_a)} \nonumber \\
&\times \exp \left[ -3 w_a \frac{z}{1+z} \right] \;. \nonumber
\end{align}
Dominant error contribution to $d_L$ is the lensing error, explicitly given in Eq.~(\ref{eq19}). After substituting $\sigma_{\rm{lens}}$ for $\Delta d_L/d_L$, we plot numerically evaluating $\Delta X(z)$ in Fig.~\ref{fig9}, assuming parameters $r_n=1$, $T_{\rm{obs}}=3\,{\rm{yr}}$, and $\dot{n}_0=10^{-6}\,{\rm{Mpc}}^{-3}\,{\rm{yr}}^{-1}$, and compare with $\Delta X(z)$ coming from instrumental noise in a GW observation. We found that the error of $d_L$ is much smaller than that of $X$ and can be neglected. This conclusion does not change if other parameters for $r_n$, $T_{\rm{obs}}$, and $\dot{n}_0$ are used.

\bibliography{X-GWB-NS}

\begin{thebibliography}{90}
\expandafter\ifx\csname natexlab\endcsname\relax\def\natexlab#1{#1}\fi
\expandafter\ifx\csname bibnamefont\endcsname\relax
  \def\bibnamefont#1{#1}\fi
\expandafter\ifx\csname bibfnamefont\endcsname\relax
  \def\bibfnamefont#1{#1}\fi
\expandafter\ifx\csname citenamefont\endcsname\relax
  \def\citenamefont#1{#1}\fi
\expandafter\ifx\csname url\endcsname\relax
  \def\url#1{\texttt{#1}}\fi
\expandafter\ifx\csname urlprefix\endcsname\relax\def\urlprefix{URL }\fi
\providecommand{\bibinfo}[2]{#2}
\providecommand{\eprint}[2][]{\url{#2}}

\bibitem[{\citenamefont{Seto et~al.}(2001)\citenamefont{Seto, Kawamura, and
  Nakamura}}]{bib1}
\bibinfo{author}{\bibfnamefont{N.}~\bibnamefont{Seto}},
  \bibinfo{author}{\bibfnamefont{S.}~\bibnamefont{Kawamura}}, \bibnamefont{and}
  \bibinfo{author}{\bibfnamefont{T.}~\bibnamefont{Nakamura}},
  \bibinfo{journal}{{Phys. Rev. Lett.}} \textbf{\bibinfo{volume}{87}},
  \bibinfo{pages}{221103} (\bibinfo{year}{2001}).

\bibitem[{\citenamefont{Kawamura et~al.}(2011)}]{bib2}
\bibinfo{author}{\bibfnamefont{S.}~\bibnamefont{Kawamura}}
  \bibnamefont{et~al.}, \bibinfo{journal}{{Classical Quantum Gravity}}
  \textbf{\bibinfo{volume}{28}}, \bibinfo{pages}{094011}
  (\bibinfo{year}{2011}).

\bibitem[{bib({\natexlab{a}})}]{bib3}
\bibinfo{note}{E. S. Phinney {\it{et al.}}, The Big Bang Observer, NASA Mission
  Concept Study (2003)}.

\bibitem[{\citenamefont{Cutler and Holz}(2009)}]{bib4}
\bibinfo{author}{\bibfnamefont{C.}~\bibnamefont{Cutler}} \bibnamefont{and}
  \bibinfo{author}{\bibfnamefont{D.~E.} \bibnamefont{Holz}},
  \bibinfo{journal}{{Phys. Rev.}} \textbf{\bibinfo{volume}{D 80}},
  \bibinfo{pages}{104009} (\bibinfo{year}{2009}).

\bibitem[{\citenamefont{Seto}(2006{\natexlab{a}})}]{bib46}
\bibinfo{author}{\bibfnamefont{N.}~\bibnamefont{Seto}},
  \bibinfo{journal}{{Phys. Rev.}} \textbf{\bibinfo{volume}{D 73}},
  \bibinfo{pages}{063001} (\bibinfo{year}{2006}{\natexlab{a}}).

\bibitem[{\citenamefont{Boyle and Buonanno}(2008)}]{bib70}
\bibinfo{author}{\bibfnamefont{L.~A.} \bibnamefont{Boyle}} \bibnamefont{and}
  \bibinfo{author}{\bibfnamefont{A.}~\bibnamefont{Buonanno}},
  \bibinfo{journal}{{Phys. Rev.}} \textbf{\bibinfo{volume}{D 78}},
  \bibinfo{pages}{043531} (\bibinfo{year}{2008}).

\bibitem[{\citenamefont{Kuroyanagi et~al.}(2010)\citenamefont{Kuroyanagi,
  Gordon, Silk, and Sugiyama}}]{bib67}
\bibinfo{author}{\bibfnamefont{S.}~\bibnamefont{Kuroyanagi}},
  \bibinfo{author}{\bibfnamefont{C.}~\bibnamefont{Gordon}},
  \bibinfo{author}{\bibfnamefont{J.}~\bibnamefont{Silk}}, \bibnamefont{and}
  \bibinfo{author}{\bibfnamefont{N.}~\bibnamefont{Sugiyama}},
  \bibinfo{journal}{{Phys. Rev.}} \textbf{\bibinfo{volume}{D 81}},
  \bibinfo{pages}{083524} (\bibinfo{year}{2010}).

\bibitem[{\citenamefont{Nakayama et~al.}(2008)\citenamefont{Nakayama, Saito,
  Suwa, and Yokoyama}}]{bib68}
\bibinfo{author}{\bibfnamefont{K.}~\bibnamefont{Nakayama}},
  \bibinfo{author}{\bibfnamefont{S.}~\bibnamefont{Saito}},
  \bibinfo{author}{\bibfnamefont{Y.}~\bibnamefont{Suwa}}, \bibnamefont{and}
  \bibinfo{author}{\bibfnamefont{J.}~\bibnamefont{Yokoyama}},
  \bibinfo{journal}{{J. Cosmol. Astropart. Phys.}}
  \textbf{\bibinfo{volume}{06}}, \bibinfo{pages}{020} (\bibinfo{year}{2008}).

\bibitem[{\citenamefont{Saito and Yokoyama}(2009)}]{bib69}
\bibinfo{author}{\bibfnamefont{R.}~\bibnamefont{Saito}} \bibnamefont{and}
  \bibinfo{author}{\bibfnamefont{J.}~\bibnamefont{Yokoyama}},
  \bibinfo{journal}{{Phys. Rev. Lett.}} \textbf{\bibinfo{volume}{102}},
  \bibinfo{pages}{161101} (\bibinfo{year}{2009}).

\bibitem[{\citenamefont{Nishizawa et~al.}(2011)\citenamefont{Nishizawa, Taruya,
  and Saito}}]{bib37}
\bibinfo{author}{\bibfnamefont{A.}~\bibnamefont{Nishizawa}},
  \bibinfo{author}{\bibfnamefont{A.}~\bibnamefont{Taruya}}, \bibnamefont{and}
  \bibinfo{author}{\bibfnamefont{S.}~\bibnamefont{Saito}},
  \bibinfo{journal}{{Phys. Rev.}} \textbf{\bibinfo{volume}{D 83}},
  \bibinfo{pages}{084045} (\bibinfo{year}{2011}).

\bibitem[{\citenamefont{Gair et~al.}(2009)\citenamefont{Gair, Mandel, Sesana,
  and Vecchio}}]{bib64}
\bibinfo{author}{\bibfnamefont{J.~R.} \bibnamefont{Gair}},
  \bibinfo{author}{\bibfnamefont{I.}~\bibnamefont{Mandel}},
  \bibinfo{author}{\bibfnamefont{A.}~\bibnamefont{Sesana}}, \bibnamefont{and}
  \bibinfo{author}{\bibfnamefont{A.}~\bibnamefont{Vecchio}},
  \bibinfo{journal}{{Classical Quantum Gravity}} \textbf{\bibinfo{volume}{26}},
  \bibinfo{pages}{204009} (\bibinfo{year}{2009}).

\bibitem[{\citenamefont{Yagi and Tanaka}(2010)}]{bib5}
\bibinfo{author}{\bibfnamefont{K.}~\bibnamefont{Yagi}} \bibnamefont{and}
  \bibinfo{author}{\bibfnamefont{T.}~\bibnamefont{Tanaka}},
  \bibinfo{journal}{{Prog. Theor. Phys.}} \textbf{\bibinfo{volume}{123}},
  \bibinfo{pages}{1069} (\bibinfo{year}{2010}).

\bibitem[{\citenamefont{Seto}(2006{\natexlab{b}})}]{bib65}
\bibinfo{author}{\bibfnamefont{N.}~\bibnamefont{Seto}},
  \bibinfo{journal}{{Phys. Rev. Lett.}} \textbf{\bibinfo{volume}{97}},
  \bibinfo{pages}{151101} (\bibinfo{year}{2006}{\natexlab{b}}).

\bibitem[{\citenamefont{Seto}(2007)}]{bib66}
\bibinfo{author}{\bibfnamefont{N.}~\bibnamefont{Seto}},
  \bibinfo{journal}{{Phys. Rev.}} \textbf{\bibinfo{volume}{D 75}},
  \bibinfo{pages}{061302(R)} (\bibinfo{year}{2007}).

\bibitem[{\citenamefont{Nishizawa et~al.}(2010)\citenamefont{Nishizawa, Taruya,
  and Kawamura}}]{bib48}
\bibinfo{author}{\bibfnamefont{A.}~\bibnamefont{Nishizawa}},
  \bibinfo{author}{\bibfnamefont{A.}~\bibnamefont{Taruya}}, \bibnamefont{and}
  \bibinfo{author}{\bibfnamefont{S.}~\bibnamefont{Kawamura}},
  \bibinfo{journal}{{Phys. Rev.}} \textbf{\bibinfo{volume}{D 81}},
  \bibinfo{pages}{104043} (\bibinfo{year}{2010}).

\bibitem[{\citenamefont{Schutz}(1986)}]{bib6}
\bibinfo{author}{\bibfnamefont{B.~F.} \bibnamefont{Schutz}},
  \bibinfo{journal}{{Nature (London)}} \textbf{\bibinfo{volume}{323}},
  \bibinfo{pages}{310} (\bibinfo{year}{1986}).

\bibitem[{\citenamefont{Holz and Hughes}(2005)}]{bib21}
\bibinfo{author}{\bibfnamefont{D.~E.} \bibnamefont{Holz}} \bibnamefont{and}
  \bibinfo{author}{\bibfnamefont{S.~A.} \bibnamefont{Hughes}},
  \bibinfo{journal}{{Astrophys. J.}} \textbf{\bibinfo{volume}{629}},
  \bibinfo{pages}{15} (\bibinfo{year}{2005}).

\bibitem[{\citenamefont{Dalal et~al.}(2006)\citenamefont{Dalal, Holz, Hughes,
  and Jain}}]{bib22}
\bibinfo{author}{\bibfnamefont{N.}~\bibnamefont{Dalal}},
  \bibinfo{author}{\bibfnamefont{D.~E.} \bibnamefont{Holz}},
  \bibinfo{author}{\bibfnamefont{S.~A.} \bibnamefont{Hughes}},
  \bibnamefont{and} \bibinfo{author}{\bibfnamefont{B.}~\bibnamefont{Jain}},
  \bibinfo{journal}{{Phys. Rev.}} \textbf{\bibinfo{volume}{D 74}},
  \bibinfo{pages}{063006} (\bibinfo{year}{2006}).

\bibitem[{\citenamefont{Arun et~al.}(2007)\citenamefont{Arun, Iyer,
  Sathyaprakash, Sinha, and Broeck}}]{bib23}
\bibinfo{author}{\bibfnamefont{K.~G.} \bibnamefont{Arun}},
  \bibinfo{author}{\bibfnamefont{B.~R.} \bibnamefont{Iyer}},
  \bibinfo{author}{\bibfnamefont{B.~S.} \bibnamefont{Sathyaprakash}},
  \bibinfo{author}{\bibfnamefont{S.}~\bibnamefont{Sinha}}, \bibnamefont{and}
  \bibinfo{author}{\bibfnamefont{C.~V.~D.} \bibnamefont{Broeck}},
  \bibinfo{journal}{{Phys. Rev.}} \textbf{\bibinfo{volume}{D 76}},
  \bibinfo{pages}{104016} (\bibinfo{year}{2007}).

\bibitem[{\citenamefont{Deffayet and Menou}(2007)}]{bib24}
\bibinfo{author}{\bibfnamefont{C.}~\bibnamefont{Deffayet}} \bibnamefont{and}
  \bibinfo{author}{\bibfnamefont{K.}~\bibnamefont{Menou}},
  \bibinfo{journal}{{Astrophys. J.}} \textbf{\bibinfo{volume}{668}},
  \bibinfo{pages}{L143} (\bibinfo{year}{2007}).

\bibitem[{\citenamefont{Nissanke et~al.}(2010)\citenamefont{Nissanke, Holz,
  Hughes, Dalal, and Sievers}}]{bib25}
\bibinfo{author}{\bibfnamefont{S.}~\bibnamefont{Nissanke}},
  \bibinfo{author}{\bibfnamefont{D.~E.} \bibnamefont{Holz}},
  \bibinfo{author}{\bibfnamefont{S.~A.} \bibnamefont{Hughes}},
  \bibinfo{author}{\bibfnamefont{N.}~\bibnamefont{Dalal}}, \bibnamefont{and}
  \bibinfo{author}{\bibfnamefont{J.~L.} \bibnamefont{Sievers}},
  \bibinfo{journal}{{Astrophys. J.}} \textbf{\bibinfo{volume}{725}},
  \bibinfo{pages}{496} (\bibinfo{year}{2010}).

\bibitem[{\citenamefont{Sathyaprakash et~al.}(2010)\citenamefont{Sathyaprakash,
  Shutz, and Broeck}}]{bib26}
\bibinfo{author}{\bibfnamefont{B.~S.} \bibnamefont{Sathyaprakash}},
  \bibinfo{author}{\bibfnamefont{B.~F.} \bibnamefont{Shutz}}, \bibnamefont{and}
  \bibinfo{author}{\bibfnamefont{C.~V.~D.} \bibnamefont{Broeck}},
  \bibinfo{journal}{{Classical Quantum Gravity}} \textbf{\bibinfo{volume}{27}},
  \bibinfo{pages}{215006} (\bibinfo{year}{2010}).

\bibitem[{\citenamefont{Broeck et~al.}(2010)\citenamefont{Broeck, Trias,
  Sathyaprakash, and Sintes}}]{bib27}
\bibinfo{author}{\bibfnamefont{C.~V.~D.} \bibnamefont{Broeck}},
  \bibinfo{author}{\bibfnamefont{M.}~\bibnamefont{Trias}},
  \bibinfo{author}{\bibfnamefont{B.~S.} \bibnamefont{Sathyaprakash}},
  \bibnamefont{and} \bibinfo{author}{\bibfnamefont{A.~M.}
  \bibnamefont{Sintes}}, \bibinfo{journal}{{Phys. Rev.}}
  \textbf{\bibinfo{volume}{D 81}}, \bibinfo{pages}{124031}
  (\bibinfo{year}{2010}).

\bibitem[{\citenamefont{Zhao et~al.}(2011)\citenamefont{Zhao, Broeck, Baskaran,
  and Li}}]{bib28}
\bibinfo{author}{\bibfnamefont{W.}~\bibnamefont{Zhao}},
  \bibinfo{author}{\bibfnamefont{C.~V.~D.} \bibnamefont{Broeck}},
  \bibinfo{author}{\bibfnamefont{D.}~\bibnamefont{Baskaran}}, \bibnamefont{and}
  \bibinfo{author}{\bibfnamefont{T.~G.~F.} \bibnamefont{Li}},
  \bibinfo{journal}{{Phys. Rev.}} \textbf{\bibinfo{volume}{D 83}},
  \bibinfo{pages}{023005} (\bibinfo{year}{2011}).

\bibitem[{\citenamefont{Petiteau et~al.}(2011)\citenamefont{Petiteau, Babak,
  and Sesana}}]{bib88}
\bibinfo{author}{\bibfnamefont{A.}~\bibnamefont{Petiteau}},
  \bibinfo{author}{\bibfnamefont{S.}~\bibnamefont{Babak}}, \bibnamefont{and}
  \bibinfo{author}{\bibfnamefont{A.}~\bibnamefont{Sesana}},
  \bibinfo{journal}{Astrophys. J.} \textbf{\bibinfo{volume}{732}},
  \bibinfo{pages}{82} (\bibinfo{year}{2011}).

\bibitem[{\citenamefont{Wang et~al.}(2010)}]{bib84}
\bibinfo{author}{\bibfnamefont{Y.}~\bibnamefont{Wang}} \bibnamefont{et~al.},
  \bibinfo{journal}{{Mon. Not. R. Astron. Soc.}}
  \textbf{\bibinfo{volume}{409}}, \bibinfo{pages}{737} (\bibinfo{year}{2010}).

\bibitem[{\citenamefont{Takahashi and Nakamura}(2005)}]{bib20}
\bibinfo{author}{\bibfnamefont{R.}~\bibnamefont{Takahashi}} \bibnamefont{and}
  \bibinfo{author}{\bibfnamefont{T.}~\bibnamefont{Nakamura}},
  \bibinfo{journal}{{Prog. Theor. Phys.}} \textbf{\bibinfo{volume}{113}},
  \bibinfo{pages}{63} (\bibinfo{year}{2005}).

\bibitem[{\citenamefont{Cutler and Harms}(2006)}]{bib15}
\bibinfo{author}{\bibfnamefont{C.}~\bibnamefont{Cutler}} \bibnamefont{and}
  \bibinfo{author}{\bibfnamefont{J.}~\bibnamefont{Harms}},
  \bibinfo{journal}{{Phys. Rev.}} \textbf{\bibinfo{volume}{D 73}},
  \bibinfo{pages}{042001} (\bibinfo{year}{2006}).

\bibitem[{\citenamefont{Yagi and Seto}(2011)}]{bib17}
\bibinfo{author}{\bibfnamefont{K.}~\bibnamefont{Yagi}} \bibnamefont{and}
  \bibinfo{author}{\bibfnamefont{N.}~\bibnamefont{Seto}},
  \bibinfo{journal}{{Phys. Rev.}} \textbf{\bibinfo{volume}{D 83}},
  \bibinfo{pages}{044011} (\bibinfo{year}{2011}).

\bibitem[{bib({\natexlab{b}})}]{bib11}
\bibinfo{note}{DECIGO design parameters: optical-cavity arm length
  $1000\,{\rm{km}}$, laser power per arm $10\,{\rm{W}}$, laser wavelength
  $532\,{\rm{nm}}$, mirror diameter $1\,{\rm{m}}$, mirror mass
  $100\,{\rm{kg}}$, arm opening angle $60^{\circ}$, finesse $10$.}

\bibitem[{\citenamefont{Farmer and Phinney}(2003)}]{bib7}
\bibinfo{author}{\bibfnamefont{A.~J.} \bibnamefont{Farmer}} \bibnamefont{and}
  \bibinfo{author}{\bibfnamefont{E.~S.} \bibnamefont{Phinney}},
  \bibinfo{journal}{{Mon. Not. R. Astron. Soc.}}
  \textbf{\bibinfo{volume}{346}}, \bibinfo{pages}{1197} (\bibinfo{year}{2003}).

\bibitem[{\citenamefont{Barack and Cutler}(2004)}]{bib12}
\bibinfo{author}{\bibfnamefont{L.}~\bibnamefont{Barack}} \bibnamefont{and}
  \bibinfo{author}{\bibfnamefont{C.}~\bibnamefont{Cutler}},
  \bibinfo{journal}{{Phys. Rev.}} \textbf{\bibinfo{volume}{D 69}},
  \bibinfo{pages}{082005} (\bibinfo{year}{2004}).

\bibitem[{bib({\natexlab{c}})}]{bib14}
\bibinfo{note}{E. S. Phinney, arXiv:astro-ph/0108028 (2001)}.

\bibitem[{\citenamefont{Schneider et~al.}(2001)\citenamefont{Schneider,
  Ferrari, Matarrese, and Zwart}}]{bib16}
\bibinfo{author}{\bibfnamefont{R.}~\bibnamefont{Schneider}},
  \bibinfo{author}{\bibfnamefont{V.}~\bibnamefont{Ferrari}},
  \bibinfo{author}{\bibfnamefont{S.}~\bibnamefont{Matarrese}},
  \bibnamefont{and} \bibinfo{author}{\bibfnamefont{S.~F.~P.}
  \bibnamefont{Zwart}}, \bibinfo{journal}{{Mon. Not. R. Astron. Soc.}}
  \textbf{\bibinfo{volume}{324}}, \bibinfo{pages}{797} (\bibinfo{year}{2001}).

\bibitem[{\citenamefont{Cutler and Flanagan}(1994)}]{bib18}
\bibinfo{author}{\bibfnamefont{C.}~\bibnamefont{Cutler}} \bibnamefont{and}
  \bibinfo{author}{\bibfnamefont{E.~E.} \bibnamefont{Flanagan}},
  \bibinfo{journal}{{Phys. Rev.}} \textbf{\bibinfo{volume}{D 49}},
  \bibinfo{pages}{2658} (\bibinfo{year}{1994}).

\bibitem[{bib({\natexlab{d}})}]{bib19}
\bibinfo{note}{M. Maggiore, {\it{Gravitational Waves}} (Oxford university
  press, 2008)}.

\bibitem[{bib({\natexlab{e}})}]{bib80}
\bibinfo{note}{For the geometric factor arising from the nonorthogonal detector
  arms ($60^\circ$), we incorporate this effect into the detector noise curve
  $S_{\rm{h}}^{\rm{inst}}(f)$. The factor arising from detector's angular
  response is also taken into account in evaluating the noise curve by
  averaging over the sky.}

\bibitem[{bib({\natexlab{f}})}]{bib91}
\bibinfo{note}{There are proposals that the degeneracy could be broken if we
  {\it{a priori}} know the equation of state of a NS \cite{bib89} or if the
  observed chirp mass distribution of NSs is narrow enough \cite{bib92}.}

\bibitem[{\citenamefont{Fairhurst}(2011)}]{bib77}
\bibinfo{author}{\bibfnamefont{S.}~\bibnamefont{Fairhurst}},
  \bibinfo{journal}{{Classical Quantum Gravity}} \textbf{\bibinfo{volume}{28}},
  \bibinfo{pages}{105021} (\bibinfo{year}{2011}).

\bibitem[{\citenamefont{Nuttall and Sutton}(2010)}]{bib78}
\bibinfo{author}{\bibfnamefont{L.~K.} \bibnamefont{Nuttall}} \bibnamefont{and}
  \bibinfo{author}{\bibfnamefont{P.~J.} \bibnamefont{Sutton}},
  \bibinfo{journal}{{Phys. Rev.}} \textbf{\bibinfo{volume}{D 82}},
  \bibinfo{pages}{102002} (\bibinfo{year}{2010}).

\bibitem[{\citenamefont{Nissanke et~al.}(2011)\citenamefont{Nissanke, Sievers,
  Dalal, and Holz}}]{bib79}
\bibinfo{author}{\bibfnamefont{S.~M.} \bibnamefont{Nissanke}},
  \bibinfo{author}{\bibfnamefont{J.~L.} \bibnamefont{Sievers}},
  \bibinfo{author}{\bibfnamefont{N.}~\bibnamefont{Dalal}}, \bibnamefont{and}
  \bibinfo{author}{\bibfnamefont{D.~E.} \bibnamefont{Holz}},
  \bibinfo{journal}{Astrophys. J.} \textbf{\bibinfo{volume}{739}},
  \bibinfo{pages}{99} (\bibinfo{year}{2011}).

\bibitem[{bib({\natexlab{g}})}]{bib58}
\bibinfo{note}{G. M. Harry et al., Classical Quantum Gravity 27, 084006 (2010);
  Advanced LIGO webpage, http://www.advancedligo.mit.edu/}.

\bibitem[{bib({\natexlab{h}})}]{bib59}
\bibinfo{note}{Advanced VIRGO webpage,
  https://wwwcascina.virgo.infn.it/advirgo/}.

\bibitem[{bib({\natexlab{i}})}]{bib60}
\bibinfo{note}{K. Kuroda et al., Classical Quantum Gravity 27, 084004 (2010);
  LCGT webpage, http://gwcenter.icrr.u-tokyo.ac.jp/en/}.

\bibitem[{bib({\natexlab{j}})}]{bib61}
\bibinfo{note}{D. G. Blair et al., {J. Phys. Conf. Ser.} {\bf{122}}, 012001
  (2008); AIGRC webpage, http://www.gravity.uwa.edu.au/}.

\bibitem[{\citenamefont{Chevallier and Polarski}(2001)}]{bib93}
\bibinfo{author}{\bibfnamefont{M.}~\bibnamefont{Chevallier}} \bibnamefont{and}
  \bibinfo{author}{\bibfnamefont{D.}~\bibnamefont{Polarski}},
  \bibinfo{journal}{{Int. J. Mod. Phys.}} \textbf{\bibinfo{volume}{D 10}},
  \bibinfo{pages}{213} (\bibinfo{year}{2001}).

\bibitem[{\citenamefont{Finn}(1992)}]{bib29}
\bibinfo{author}{\bibfnamefont{L.~S.} \bibnamefont{Finn}},
  \bibinfo{journal}{{Phys. Rev.}} \textbf{\bibinfo{volume}{D 46}},
  \bibinfo{pages}{5236} (\bibinfo{year}{1992}).

\bibitem[{\citenamefont{Riess et~al.}(2011)}]{bib32}
\bibinfo{author}{\bibfnamefont{A.~G.} \bibnamefont{Riess}}
  \bibnamefont{et~al.}, \bibinfo{journal}{{Astrophys. J.}}
  \textbf{\bibinfo{volume}{730}}, \bibinfo{pages}{119} (\bibinfo{year}{2011}).

\bibitem[{\citenamefont{Abadie et~al.}(2010)}]{bib30}
\bibinfo{author}{\bibfnamefont{J.}~\bibnamefont{Abadie}} \bibnamefont{et~al.},
  \bibinfo{journal}{{Classical Quantum Gravity}} \textbf{\bibinfo{volume}{27}},
  \bibinfo{pages}{173001} (\bibinfo{year}{2010}).

\bibitem[{bib({\natexlab{k}})}]{bib71}
\bibinfo{note}{A. Albrecht et al., arXiv:astro-ph/0609591 (2006)}.

\bibitem[{\citenamefont{Beckwith et~al.}(2006)}]{bib85}
\bibinfo{author}{\bibfnamefont{S.~V.~W.} \bibnamefont{Beckwith}}
  \bibnamefont{et~al.}, \bibinfo{journal}{{Astronomical J.}}
  \textbf{\bibinfo{volume}{132}}, \bibinfo{pages}{1729} (\bibinfo{year}{2006}).

\bibitem[{bib({\natexlab{l}})}]{bib86}
\bibinfo{note}{JDEM: http://jdem.gsfc.nasa.gov/.}

\bibitem[{bib({\natexlab{m}})}]{bib87}
\bibinfo{note}{Eucrid: http://sci.esa.int/euclid.}

\bibitem[{\citenamefont{Racusin et~al.}(2011)}]{bib96}
\bibinfo{author}{\bibfnamefont{J.~L.} \bibnamefont{Racusin}}
  \bibnamefont{et~al.}, \bibinfo{journal}{{Astrophys. J.}}
  \textbf{\bibinfo{volume}{738}}, \bibinfo{pages}{138} (\bibinfo{year}{2011}).

\bibitem[{\citenamefont{Allen and Romano}(1999)}]{bib44}
\bibinfo{author}{\bibfnamefont{B.}~\bibnamefont{Allen}} \bibnamefont{and}
  \bibinfo{author}{\bibfnamefont{J.~D.} \bibnamefont{Romano}},
  \bibinfo{journal}{{Phys. Rev.}} \textbf{\bibinfo{volume}{D 59}},
  \bibinfo{pages}{102001} (\bibinfo{year}{1999}).

\bibitem[{\citenamefont{Christensen}(1992)}]{bib42}
\bibinfo{author}{\bibfnamefont{N.}~\bibnamefont{Christensen}},
  \bibinfo{journal}{{Phys. Rev.}} \textbf{\bibinfo{volume}{D 46}},
  \bibinfo{pages}{5250} (\bibinfo{year}{1992}).

\bibitem[{\citenamefont{Flanagan}(1993)}]{bib43}
\bibinfo{author}{\bibfnamefont{E.~E.} \bibnamefont{Flanagan}},
  \bibinfo{journal}{{Phys. Rev.}} \textbf{\bibinfo{volume}{D 48}},
  \bibinfo{pages}{2389} (\bibinfo{year}{1993}).

\bibitem[{\citenamefont{Corbin and Cornish}(2006)}]{bib45}
\bibinfo{author}{\bibfnamefont{V.}~\bibnamefont{Corbin}} \bibnamefont{and}
  \bibinfo{author}{\bibfnamefont{N.~J.} \bibnamefont{Cornish}},
  \bibinfo{journal}{{Classical Quantum Gravity}} \textbf{\bibinfo{volume}{23}},
  \bibinfo{pages}{2446} (\bibinfo{year}{2006}).

\bibitem[{\citenamefont{Kudoh et~al.}(2006)\citenamefont{Kudoh, Taruya,
  Hiramatsu, and Himemoto}}]{bib47}
\bibinfo{author}{\bibfnamefont{H.}~\bibnamefont{Kudoh}},
  \bibinfo{author}{\bibfnamefont{A.}~\bibnamefont{Taruya}},
  \bibinfo{author}{\bibfnamefont{T.}~\bibnamefont{Hiramatsu}},
  \bibnamefont{and} \bibinfo{author}{\bibfnamefont{Y.}~\bibnamefont{Himemoto}},
  \bibinfo{journal}{{Phys. Rev.}} \textbf{\bibinfo{volume}{D 73}},
  \bibinfo{pages}{064006} (\bibinfo{year}{2006}).

\bibitem[{bib({\natexlab{n}})}]{bib82}
\bibinfo{note}{In the observational frequency band of DECIGO, the overlap
  reduction function can be approximately set to unity. On the other hand, this
  is not true for BBO, because BBO is a transponder type like LISA and the
  response to a GW is degraded more and more at high frequencies. However, the
  overlap reduction function also can also be approximated to unity at low
  frequencies below $1\,{\rm{Hz}}$, where the GW signal largely contributes to
  the SNR frequency integral.}

\bibitem[{\citenamefont{Prince et~al.}(2002)\citenamefont{Prince, Tinto,
  Larson, and Armstrong}}]{bib49}
\bibinfo{author}{\bibfnamefont{T.~A.} \bibnamefont{Prince}},
  \bibinfo{author}{\bibfnamefont{M.}~\bibnamefont{Tinto}},
  \bibinfo{author}{\bibfnamefont{S.~L.} \bibnamefont{Larson}},
  \bibnamefont{and} \bibinfo{author}{\bibfnamefont{J.~W.}
  \bibnamefont{Armstrong}}, \bibinfo{journal}{{Phys. Rev.}}
  \textbf{\bibinfo{volume}{D 66}}, \bibinfo{pages}{122002}
  (\bibinfo{year}{2002}).

\bibitem[{bib({\natexlab{o}})}]{bib51}
\bibinfo{note}{D. Samtleben et al., arXiv:0806.4334 (2008)}.

\bibitem[{bib({\natexlab{p}})}]{bib52}
\bibinfo{note}{C. E. North et al., arXiv:0805.3690 (2008)}.

\bibitem[{\citenamefont{Crill et~al.}(2008)}]{bib53}
\bibinfo{author}{\bibfnamefont{B.~P.} \bibnamefont{Crill}}
  \bibnamefont{et~al.}, \bibinfo{journal}{SPIE}
  \textbf{\bibinfo{volume}{7010}}, \bibinfo{pages}{79} (\bibinfo{year}{2008}).

\bibitem[{\citenamefont{Oxley et~al.}(2004)}]{bib54}
\bibinfo{author}{\bibfnamefont{P.}~\bibnamefont{Oxley}} \bibnamefont{et~al.},
  \bibinfo{journal}{SPIE} \textbf{\bibinfo{volume}{5543}}, \bibinfo{pages}{320}
  (\bibinfo{year}{2004}).

\bibitem[{\citenamefont{Liddle}(1994)}]{bib55}
\bibinfo{author}{\bibfnamefont{A.~R.} \bibnamefont{Liddle}},
  \bibinfo{journal}{{Phys. Rev.}} \textbf{\bibinfo{volume}{D 49}},
  \bibinfo{pages}{3805} (\bibinfo{year}{1994}).

\bibitem[{\citenamefont{Turner et~al.}(1993)\citenamefont{Turner, White, and
  Lidsey}}]{bib56}
\bibinfo{author}{\bibfnamefont{M.~S.} \bibnamefont{Turner}},
  \bibinfo{author}{\bibfnamefont{M.}~\bibnamefont{White}}, \bibnamefont{and}
  \bibinfo{author}{\bibfnamefont{J.~E.} \bibnamefont{Lidsey}},
  \bibinfo{journal}{{Phys. Rev.}} \textbf{\bibinfo{volume}{D 48}},
  \bibinfo{pages}{4613} (\bibinfo{year}{1993}).

\bibitem[{\citenamefont{Turner}(1997)}]{bib57}
\bibinfo{author}{\bibfnamefont{M.~S.} \bibnamefont{Turner}},
  \bibinfo{journal}{{Phys. Rev.}} \textbf{\bibinfo{volume}{D 55}},
  \bibinfo{pages}{R435} (\bibinfo{year}{1997}).

\bibitem[{\citenamefont{Chongchitnan and Efstathiou}(2006)}]{bib62}
\bibinfo{author}{\bibfnamefont{S.}~\bibnamefont{Chongchitnan}}
  \bibnamefont{and}
  \bibinfo{author}{\bibfnamefont{G.}~\bibnamefont{Efstathiou}},
  \bibinfo{journal}{{Phys. Rev.}} \textbf{\bibinfo{volume}{D 73}},
  \bibinfo{pages}{083511} (\bibinfo{year}{2006}).

\bibitem[{\citenamefont{Friedman et~al.}(2006)\citenamefont{Friedman, Cooray,
  and Melchiorri}}]{bib63}
\bibinfo{author}{\bibfnamefont{B.~C.} \bibnamefont{Friedman}},
  \bibinfo{author}{\bibfnamefont{A.}~\bibnamefont{Cooray}}, \bibnamefont{and}
  \bibinfo{author}{\bibfnamefont{A.}~\bibnamefont{Melchiorri}},
  \bibinfo{journal}{{Phys. Rev.}} \textbf{\bibinfo{volume}{D 74}},
  \bibinfo{pages}{123509} (\bibinfo{year}{2006}).

\bibitem[{bib({\natexlab{q}})}]{bib83}
\bibinfo{note}{There are some inflationary models that have a blue-tilted
  spectrum with a peak at high frequencies, e.g. quintessential inflation
  \cite{bib75,bib76}. There are also some other mechanisms generating a large
  GW amplitude at high frequencies such as preheating, phase transition, and
  comological defects.}

\bibitem[{\citenamefont{Komatsu et~al.}(2011)}]{bib50}
\bibinfo{author}{\bibfnamefont{E.}~\bibnamefont{Komatsu}} \bibnamefont{et~al.},
  \bibinfo{journal}{Astrophys. J. Suppl. Ser.} \textbf{\bibinfo{volume}{192}},
  \bibinfo{pages}{18} (\bibinfo{year}{2011}).

\bibitem[{\citenamefont{Uzan et~al.}(2008)\citenamefont{Uzan, Bernardeau, and
  Mellier}}]{bib72}
\bibinfo{author}{\bibfnamefont{J.~P.} \bibnamefont{Uzan}},
  \bibinfo{author}{\bibfnamefont{F.}~\bibnamefont{Bernardeau}},
  \bibnamefont{and} \bibinfo{author}{\bibfnamefont{Y.}~\bibnamefont{Mellier}},
  \bibinfo{journal}{{Phys. Rev.}} \textbf{\bibinfo{volume}{D 77}},
  \bibinfo{pages}{021301(R)} (\bibinfo{year}{2008}).

\bibitem[{\citenamefont{Amendola et~al.}(2008)\citenamefont{Amendola, Balbi,
  and Quercellini}}]{bib73}
\bibinfo{author}{\bibfnamefont{L.}~\bibnamefont{Amendola}},
  \bibinfo{author}{\bibfnamefont{A.}~\bibnamefont{Balbi}}, \bibnamefont{and}
  \bibinfo{author}{\bibfnamefont{C.}~\bibnamefont{Quercellini}},
  \bibinfo{journal}{{Phys. Lett.}} \textbf{\bibinfo{volume}{B 660}},
  \bibinfo{pages}{81} (\bibinfo{year}{2008}).

\bibitem[{\citenamefont{Quercellini et~al.}(2008)\citenamefont{Quercellini,
  Amendola, and Balbi}}]{bib74}
\bibinfo{author}{\bibfnamefont{C.}~\bibnamefont{Quercellini}},
  \bibinfo{author}{\bibfnamefont{L.}~\bibnamefont{Amendola}}, \bibnamefont{and}
  \bibinfo{author}{\bibfnamefont{A.}~\bibnamefont{Balbi}},
  \bibinfo{journal}{{Mon. Not. R. Astron. Soc.}}
  \textbf{\bibinfo{volume}{391}}, \bibinfo{pages}{1308} (\bibinfo{year}{2008}).

\bibitem[{\citenamefont{Harms et~al.}(2008)\citenamefont{Harms, Mahrdt, Otto,
  and Prie$\beta$}}]{bib90}
\bibinfo{author}{\bibfnamefont{J.}~\bibnamefont{Harms}},
  \bibinfo{author}{\bibfnamefont{C.}~\bibnamefont{Mahrdt}},
  \bibinfo{author}{\bibfnamefont{M.}~\bibnamefont{Otto}}, \bibnamefont{and}
  \bibinfo{author}{\bibfnamefont{M.}~\bibnamefont{Prie$\beta$}},
  \bibinfo{journal}{{Phys. Rev.}} \textbf{\bibinfo{volume}{D 77}},
  \bibinfo{pages}{123010} (\bibinfo{year}{2008}).

\bibitem[{bib({\natexlab{r}})}]{bib89}
\bibinfo{note}{C. Messenger and J. Read, arXiv:1107.5725 (2011)}.

\bibitem[{bib({\natexlab{s}})}]{bib92}
\bibinfo{note}{S. R. Taylor, J. R. Gair, and I. Mandel, arXiv:1108.5161
  (2011)}.

\bibitem[{\citenamefont{Punturo et~al.}(2010)}]{bib94}
\bibinfo{author}{\bibfnamefont{M.}~\bibnamefont{Punturo}} \bibnamefont{et~al.},
  \bibinfo{journal}{{Classical Quantum Gravity}} \textbf{\bibinfo{volume}{27}},
  \bibinfo{pages}{084007} (\bibinfo{year}{2010}).

\bibitem[{bib({\natexlab{t}})}]{bib95}
\bibinfo{note}{G. M. Harry (for the LIGO Scientific Collaboration), Classical
  Quantum Gravity 27, 084006 (2010).}

\bibitem[{bib({\natexlab{u}})}]{bib81}
\bibinfo{note}{Yagi and Seto set the lower frequency cutoff as
  $f_{\rm{min}}=0.2\,{\rm{Hz}}$. However, lower $f_{\rm{min}}$ that we use in
  this paper does not significantly affects the SNR, because the SNR is
  degraded below the frequency $0.2\,{\rm{Hz}}$ due to the existence of the WD
  foreground.}

\bibitem[{\citenamefont{Wambsganss et~al.}(1997)\citenamefont{Wambsganss, Cen,
  Xu, and Ostriker}}]{bib38}
\bibinfo{author}{\bibfnamefont{J.}~\bibnamefont{Wambsganss}},
  \bibinfo{author}{\bibfnamefont{R.}~\bibnamefont{Cen}},
  \bibinfo{author}{\bibfnamefont{G.}~\bibnamefont{Xu}}, \bibnamefont{and}
  \bibinfo{author}{\bibfnamefont{J.~P.} \bibnamefont{Ostriker}},
  \bibinfo{journal}{{Astrophys. J.}} \textbf{\bibinfo{volume}{475}},
  \bibinfo{pages}{L81} (\bibinfo{year}{1997}).

\bibitem[{\citenamefont{Holz and Wald}(1998)}]{bib39}
\bibinfo{author}{\bibfnamefont{D.~E.} \bibnamefont{Holz}} \bibnamefont{and}
  \bibinfo{author}{\bibfnamefont{R.~M.} \bibnamefont{Wald}},
  \bibinfo{journal}{{Phys. Rev.}} \textbf{\bibinfo{volume}{D 58}},
  \bibinfo{pages}{063501} (\bibinfo{year}{1998}).

\bibitem[{\citenamefont{Holz and Linder}(2005)}]{bib33}
\bibinfo{author}{\bibfnamefont{D.~E.} \bibnamefont{Holz}} \bibnamefont{and}
  \bibinfo{author}{\bibfnamefont{E.~V.} \bibnamefont{Linder}},
  \bibinfo{journal}{Astrophys. J.} \textbf{\bibinfo{volume}{631}},
  \bibinfo{pages}{678} (\bibinfo{year}{2005}).

\bibitem[{\citenamefont{Kainulainen and Marra}(2011)}]{bib40}
\bibinfo{author}{\bibfnamefont{K.}~\bibnamefont{Kainulainen}} \bibnamefont{and}
  \bibinfo{author}{\bibfnamefont{V.}~\bibnamefont{Marra}},
  \bibinfo{journal}{{Phys. Rev.}} \textbf{\bibinfo{volume}{D 83}},
  \bibinfo{pages}{023009} (\bibinfo{year}{2011}).

\bibitem[{\citenamefont{Hirata et~al.}(2010)\citenamefont{Hirata, Holz, and
  Cutler}}]{bib34}
\bibinfo{author}{\bibfnamefont{C.~M.} \bibnamefont{Hirata}},
  \bibinfo{author}{\bibfnamefont{D.~E.} \bibnamefont{Holz}}, \bibnamefont{and}
  \bibinfo{author}{\bibfnamefont{C.}~\bibnamefont{Cutler}},
  \bibinfo{journal}{{Phys. Rev.}} \textbf{\bibinfo{volume}{D 81}},
  \bibinfo{pages}{124046} (\bibinfo{year}{2010}).

\bibitem[{\citenamefont{Gordon et~al.}(2007)\citenamefont{Gordon, Land, and
  Slosar}}]{bib36}
\bibinfo{author}{\bibfnamefont{C.}~\bibnamefont{Gordon}},
  \bibinfo{author}{\bibfnamefont{K.}~\bibnamefont{Land}}, \bibnamefont{and}
  \bibinfo{author}{\bibfnamefont{A.}~\bibnamefont{Slosar}},
  \bibinfo{journal}{{Phys. Rev. Lett.}} \textbf{\bibinfo{volume}{99}},
  \bibinfo{pages}{081301} (\bibinfo{year}{2007}).

\bibitem[{\citenamefont{Silberman et~al.}(2001)\citenamefont{Silberman, Dekel,
  Eldar, and Zehavi}}]{bib41}
\bibinfo{author}{\bibfnamefont{L.}~\bibnamefont{Silberman}},
  \bibinfo{author}{\bibfnamefont{A.}~\bibnamefont{Dekel}},
  \bibinfo{author}{\bibfnamefont{A.}~\bibnamefont{Eldar}}, \bibnamefont{and}
  \bibinfo{author}{\bibfnamefont{I.}~\bibnamefont{Zehavi}},
  \bibinfo{journal}{{Astrophys. J.}} \textbf{\bibinfo{volume}{557}},
  \bibinfo{pages}{102} (\bibinfo{year}{2001}).

\bibitem[{\citenamefont{Giovannini}(1998)}]{bib75}
\bibinfo{author}{\bibfnamefont{M.}~\bibnamefont{Giovannini}},
  \bibinfo{journal}{{Phys. Rev.}} \textbf{\bibinfo{volume}{D 58}},
  \bibinfo{pages}{083504} (\bibinfo{year}{1998}).

\bibitem[{\citenamefont{Giovannini}(1999)}]{bib76}
\bibinfo{author}{\bibfnamefont{M.}~\bibnamefont{Giovannini}},
  \bibinfo{journal}{{Phys. Rev.}} \textbf{\bibinfo{volume}{D 60}},
  \bibinfo{pages}{123511} (\bibinfo{year}{1999}).

\end{thebibliography}

\end{document}